\documentclass[11pt,oneside]{article}
\usepackage[top=1in, bottom=1.2in, left=1.1in, right=1.1in]{geometry}
\usepackage{setspace}
\usepackage{amsmath, amsthm, amssymb,amsbsy,amsfonts,mathrsfs}
\usepackage{color}
\usepackage{bm}
\usepackage{natbib}
\usepackage{authblk}
\usepackage{appendix}
\usepackage{float}
\usepackage[caption = false]{subfig}
\usepackage{graphicx}
\usepackage{caption}

\usepackage{enumitem}
\usepackage{enumitem}
\usepackage[breaklinks=true]{hyperref}
\usepackage{booktabs}
\usepackage[flushleft]{threeparttable}
\usepackage{titlesec}

\csname @openrightfalse\endcsname

\setlist{nolistsep}
\makeatletter
\newcommand*{\rom}
[1]{\expandafter\@slowromancap\romannumeral #1@}
\makeatother
\DeclareMathOperator*{\argmin}{arg\min}

\DeclareMathOperator*{\diag}{diag}
\newtheorem{thm}{Theorem}

\newtheorem{ass}{Assumption}
\newtheorem{exm}{Example}
\hypersetup{
     colorlinks   = true,
     citecolor    =  blue
}
\newcommand{\Sz}{\mathsf{S}}

\begin{document}

\title{\textbf{Quantile Factor Models}\thanks{We are indebted to Ulrich M\"uller and four anonymous referees for their
constructive inputs which have greatly improved the paper. We also thank Dante Amengual, Manuel Arellano, Steve Bond, Guillaume Carlier, Valentina Corradi, Jean-Pierre Florens, Alfred Galichon, Peter R. Hansen, Jerry Hausman, Sophocles Mavroeidis, Bent Nielsen, Olivier Scaillet, Enrique Sentana, Liangjun Su, and
participants at several seminars and conferences for many helpful comments and suggestions. Financial support from the National Natural Science Foundation of China (Grant No.71703089), The Open Society Foundation, The Oxford Martin School, the Spanish Ministerio de Econom\'{\i}a y Competitividad
(grants ECO2016-78652 and Maria de Maeztu MDM 2014-0431), and MadEco-CM (grant S205/HUM-3444) is gratefully acknowledged. The usual disclaimer applies.} }
\author[1]{Liang Chen}
\author[2]{Juan J. Dolado}
\author[3]{Jes\'{u}s Gonzalo}
\affil[1]{School of Economics, Shanghai University of Finance and Economics, chen.liang@mail.shufe.edu.cn }
\affil[2]{Department of Economics, Universidad Carlos III de Madrid, dolado@eco.uc3m.es}
\affil[3]{Department of Economics, Universidad Carlos III de Madrid, jgonzalo@est-eco.uc3m.es}
\date{March 23, 2020}

\clearpage\maketitle
\thispagestyle{empty}

\begin{abstract}
Quantile Factor Models (QFM) represent a new class of factor models for high-dimensional panel data. Unlike Approximate Factor Models (AFM), where only location-shifting factors can be extracted, QFM also capture unobserved factors shifting other relevant parts of the distributions of observables. We propose a  quantile regression approach, labeled Quantile Factor Analysis (QFA), to consistently estimate all the quantile-dependent factors and loadings. Their asymptotic distribution is derived using a kernel-smoothed version of the QFA estimators. Two consistent model-selection criteria, based on information criteria and rank minimization, are developed to determine the number of factors at each quantile. Moreover, in contrast to the conditions required by Principal Components Analysis in AFM, QFA estimation remains valid even when the idiosyncratic errors have heavy-tailed distributions. Three empirical applications (regarding climate, macroeconomic and finance panel data) illustrate that extra factors shifting quantiles other than the means could be relevant for causality analysis, prediction and economic interpretation of common factors.

\vspace{0.2cm}

\end{abstract}

\affil[1]{School of Economics, Shanghai University of Finance and Economics,
chen.liang@mail.shufe.edu.cn }
\affil[2]{Department of Economics,
Universidad Carlos III de Madrid, dolado@eco.uc3m.es}
\affil[3]{Department
of Economics, Universidad Carlos III de Madrid, jgonzalo@est-eco.uc3m.es}



\newpage

\section{Introduction}

\noindent {\ }\setcounter{page}{1} \label{sectionintro} Following the key
contributions by \citet{ross1976arbitrage}, \citet{chamberlain1983arbitrage}
and \citet{connor1986performance} to the theory of approximate factor models (AFM henceforth) in the context of asset pricing, the analysis and applications of this class of models have proliferated thereafter. As it is well known, AFM imply that a panel $\{X_{it}\}$ of $N$ variables (units), each
with $T$ observations, has the representation $X_{it}=\lambda _{i}^{\prime
}f_{t}+\epsilon _{it}$, where $\lambda _{i}=[\lambda _{i1},..,\lambda _{ir}]^{\prime }$ and $f_{t}$ $=[f_{t1},..,f_{tr}]^{
\prime }$ are $r\times 1$ vectors of factor loadings and common factors, respectively, with $r\ll N$, and $\{\epsilon _{it}\}$ are zero-mean weakly dependent idiosyncratic disturbances which are uncorrelated with the factors.

The fact that it is easy to construct theories involving common factors, at least in a narrative version, together with the availability of fairly straightforward estimation procedures for AFM--- e.g. via Principal Components Analysis (PCA), has led to their extensive use in many
fields of economics.\footnote{ Early applications of AFM abound in Aggregation Theory, Consumer Theory, Business Cycle Analysis, Finance, Monetary Economics, and Monitoring and Forecasting; see,\textit{\ inter alia}, \citet{bai2003inferential}, \citet{bai2008large}, %
\citet{stock2011dynamic}. More recently, the characterization of cross-sectional dependence among error terms in Panel Data has relied on the use of a finite number of unobserved common factors which originate from economy-wide shocks affecting units with different intensities (loadings).  Interactive fixed-effects models can be easily estimated by PCA (see \citealt{bai2009panel}) or by common correlated
effects (see \citealt{pesaran2006estimation}), and there are even generalizations of these techniques for nonlinear panel single-index models (see \citealt{chen2018panel}). Lastly, the surge of Big Data and Machine Learning technologies has made factor models a key tool for dimension reduction and predictive analytics when using very large datasets (see \citealt{athey2019machine} for a survey).}

Inspired by the generalization of linear regression to quantile regression (QR) models, our starting point in this paper is to claim that the standard regression interpretation of static AFM as linear conditional mean models of $X_{it}$ given $%
f_{t}$ (i.e. $\mathbb{E}(X_{it}|f_{t})=\lambda _{i}^{\prime }f_{t}$), entails two possibly restrictive features. On the one hand, PCA does not capture hidden factors that may shift characteristics (moments or quantiles)
of the distribution of $X_{it}$ other than its mean. On the other hand, neither the loadings $\lambda _{i}$ nor the factors $f_{t}$ are allowed to vary across the distributional characteristics of each unit in the panel.

Highlighting these limitations, a growing literature in empirical finance has been documenting a much more pronounced co-movement of financial asset returns at the lower part than at the rest of their distributions. In particular, \citet{doi:10.1080/07350015.2018.1505631} reject the null hypothesis of a Gaussian copula when analyzing the cross-sectional dependence among monthly returns on individual US stocks, for which they find nonlinear tail dependence, co-skeweness and co-kurtosis. Likewise, \citet{ando2018quantile} (\textcolor{blue}{AB 2020}, hereafter) report that the common factor structures explaining the asset return distributions in global financial markets since the subprime crisis are different in the lower and the upper tails. In international finance, \citet{maravalle2018changes} find that, while global and regional factors were key in explaining fluctuations of European government bond yields between the Great Moderation and the onset of the Great Recession (lower tails of the distributions of bond yields), their role decreased afterwards, giving way to country-specific factors as the main driving forces.\footnote{The diminishing role of the former factors could result from quantitative easing and the increase in yields (upper tail of the distribution), while the emergence of the latter factors could be due to the financial fragmentation in a number of vulnerable euro-area countries during sovereign debt crisis.} On the macro side, \citet{adrian2019vulnerable} find that, while the  estimated lower conditional quantiles of the distribution of future GDP growth in the US exhibit strong dependence on current financial conditions, the upper quantiles are stable over time. Lastly, in micro theory, \citet{de2019dynamic} have recently extended the expected utility model of rational behavior to quantile utility preferences, where e.g. factor structures determining hedonic pricing of consumption goods or financial stocks may exhibit large differences across quantiles.   

A simple way of illustrating the above limitations of the conventional formulation of AFM\ is to consider the factor structure in a \textit{location-scale
shift} model with the following Data Generating Process (DGP): $%
X_{it}=\alpha _{i}f_{1t}+\eta_{i}f_{2t}\epsilon _{it}$, with $f_{1t}\neq f_{2t}$
(both are scalars), $\eta_i,f_{2t}>0$ and $\mathbb{E}(\epsilon _{it})=0$. The first factor ($f_{1t}$) shifts location, whereas the second factor ($f_{2t}$)
shifts the scale and therefore governs the volatility of shocks to $X_{it}$. This model has been proposed by \citet{herskovic2016common} to empirically document a strong co-movement of the volatilities in the idiosyncratic component of individual stock returns and firm-level cash flows.\footnote{%
This DGP is further discussed in subsection 2.2 below, where we present a larger set of illustrative models as examples of potential DGPs for $X_{it}$. Notice that the simplifying assumption of a known number of factors in this specific example is later relaxed.} Such a DGP can be rewritten in QR format
as $X_{it}=\lambda _{i}^{\prime }(\tau )f_{t}+u_{it}(\tau )$, with $0<\tau
<1 $, $\lambda _{i}(\tau )=[\alpha _{i},\eta_{i}\mathsf{Q} _{\epsilon}(\tau )]^{\prime }$, where
$\mathsf{Q} _{\epsilon}(\tau )$ represents the quantile function of $\epsilon _{it}$, $f_{t}=[f_{1t},f_{2t}]^{\prime }$, $u_{it}(\tau )=\eta_{i}f_{2t}[\epsilon _{it}-\mathsf{Q} _{\epsilon}(\tau )]$, and the conditional quantile $\mathsf{Q}_{u_{it}(\tau)}[\tau|f_{t}]=0$.%
\footnote{%
Throughout the paper we use $\mathsf{Q}_{W}[\tau |Z]$ to denote the conditional
quantile of $W$ given $Z$.} PCA will only extract the location-shifting factor $%
f_{1t}$ in this model, but it will fail to capture the scale-shifting factor $f_{2t}$ and the quantile-dependent loadings $\lambda
_{i}(\tau )$ in its QR representation. As will be explained below, our estimation procedure allows to estimate the space spanned by $f_{1t}$ and $f_{2t}$.\footnote{Given that $f_{1t}$ can be consistently estimated by PCA, it is also feasible to separate $f_{2t}$ from their joint space.}
Also notice that, when the distribution of
$\epsilon _{it}$ is symmetric, then $f_{t}$ can be considered as being quantile dependent, i.e. $f_{t}(\tau)$, since $f_{t}(\tau
)=f_{1t}$ for $\tau =0.5$, and $f_{t}(\tau
)=[f_{1t},f_{2t}]^{\prime }$ for $\tau \neq 0.5$. Together with the remaining  examples discussed below, this means that the general class of models to be considered in the sequel would be one where the loadings, factors and the number of factors are all allowed to be quantile-dependent objects,
namely, $\lambda _{i}(\tau )$, $f_{t}(\tau )$ and $r(\tau)$ for $\tau \in (0,1)$. In what follows, we coin this class of models \textit{Quantile Factor Models} (QFM, hereafter), whose detailed definition is provided in Section 2 below.

That said, our goal in this paper is to develop a common factor methodology for QFM which is flexible enough to capture the quantile-dependent objects that standard AFM tools fail to recover. To do so, we analyze their estimation and inference, including selection criteria for  the number of factors at each quantile $\tau $. Put succinctly, QFM could be thought of as capturing the same type of flexible generalization that QR techniques represent for linear regression models.


To help understand how this new methodology works, we start by proposing an estimation approach for the quantile-dependent objects in QFM, labeled
\textit{Quantile Factor Analysis} (QFA, henceforth). Our QFA estimation procedure relies on the minimization of the standard \textit{check} function in QR (instead of the conventional quadratic loss function used in AFM) to estimate jointly the common factors $f_{t}(\tau )$ and the loadings $\lambda _{i}(\tau )$ at a given quantile $\tau $, once the number of factors has been selected. However, since the objective function for QFM is not convex in the relevant parameters, we introduce an iterative QR algorithm which yields estimators of the quantile-dependent objects. We then derive their average rates of convergence, and propose two consistent selection criteria (one based on information criteria and another on rank minimization) for the number of factors at each $\tau$. In addition, we establish asymptotic normality for QFA estimators based on smoothed QR (see e.g., \citealt{horowitz1998bootstrap} and \citealt{galvao2016smoothed}). Moreover, given that QFA estimation captures all quantile-shifting factors (including those affecting the means of observed variables), our asymptotic results and the proposed selection criteria provide a natural way to differentiate AFM from QFM. 

In sum, the key contributions of this paper to the literature on factor models can be summarized as follows:
\begin{enumerate}
\item We provide a complete asymptotic analysis for a new class of factor models: QFM. In particular, we show that the average convergence rates of the QFA estimators are the same as the PCA estimators of \citet{bai2002determining} (\textcolor{blue}{BN 2002}, hereafter), which is a crucial result for proving the consistency of the two selection criteria of  the number of factors at each $\tau$. In addition, similar to \citet{bai2003inferential}, our QFA estimators based on smoothed QR are shown to converge at the parametric rates ($\sqrt{N}$ and $\sqrt{T}$) to normal distributions.
\vspace{0.2cm}
\item We argue that the problems of incidental parameters and non-smooth object functions require the use of some novel techniques in our proofs, which are borrowed from the theory of empirical processes. Moreover, our proof strategy can be easily extended to some other nonlinear factor models (e.g., probit and logit factor models considered by \citealt{chen2018panel}) with smooth object functions. Finally, as a byproduct of our approach (and in exchange for some restrictions on the dependence of the idiosyncratic errors in an AFM; see Assumption 1 below), it is shown that the QFA estimators inherit from QR certain robustness properties to the presence of outliers and heavy-tailed distributions in those error terms, which would render PCA invalid.
\item We show in the empirical section how QFA could  provide a useful tool for quantile causal analysis, density forecasting, and economic interpretation of factors by applying the proposed methodology to three different datasets related to climate, macroeconomic aggregates and stock returns.

\end{enumerate}


\bigskip \noindent\textbf{Related literature}

There is a recent literature that attempts to make the AFM setup more flexible. For example, \citet{su2017time} allow for the factor loadings to be time-varying and \citet{pelger2018interpretable} admit these loadings to be state dependent. \citet{chen2009nonlinear} provide a
theory for nonlinear PCA, where they favor sieve estimation to retrieve nonlinear factors. Finally, \citet{gorodnichenko2017level} propose an algorithm to estimate level and volatility factors simultaneously. Different from these studies, our approach to modelling nonlinearities in factor models is through the conditional quantiles of the observed data.

On top of this, there is an emerging literature on heterogeneous panel quantile models with factor structures, especially in financial economics. The main idea is that a few unobservable factors explain co-movements of
asset return distributions in a large range of asset returns observed at high frequencies, as in stock markets. In parallel and independent research, there have been two papers related to ours. First, \citet{ma2017estimation} propose estimation and inference procedures in semiparametric
quantile factor models. In these models, factor loadings/betas are smooth functions of a small number of observables under the assumption that the included factors all have non zero mean. Then, sieve techniques are used to obtain preliminary estimation of these functions for each time period. Finally, the
factor structure is imposed in a sequential fashion to estimate the factor returns by GLS under weak conditions on cross-sectional and temporal dependence. We depart from these authors in that we do not need to assume
the loadings to depend on observables and, foremost, in that not only loadings but also factors are quantile-dependent objects in our setup. Second, in a closely related paper, \textcolor{blue}{AB (2020)} use a similar setup to ours, where the unobservable factor structure is also allowed to be quantile dependent. These authors use Bayesian MCMC and frequentist estimation approaches, the latter building upon our proposed iterative procedure, as duly acknowledged in their paper. However, we differ from \textcolor{blue}{AB (2020)} in several respects which make our QFA approach valuable: (i) our assumptions are less restrictive, since we rely on properties of the density, as in QR, while \textcolor{blue}{AB (2020)} needs all the moments of the idiosyncratic errors to exist, (ii) our  proofs of the main results are different from theirs, and (iii) our rank-minimization selection criterion to estimate the number of factors is novel, behaves well in finite samples and is computationally more efficient than the information criteria-based method. 

Finally, it is noteworthy that the illustrative location-scale shift model above, where $f_{1t}$ $\neq f_{2t}$, is behind a current line of research in asset pricing which has been coined the \textquotedblleft idiosyncratic volatility
puzzle\textquotedblright\ by \citet{ang2006cross}. This approach focuses on the co-movements in the idiosyncratic volatilities of a panel of asset returns, and consists of applying PCA to (or taking cross-sectional averages of) the squared
residuals, once the mean (PCA) factors have been removed from the original variables (a procedure labeled PCA-SQ hereafter).\footnote{%
See, e.g., \citet{barigozzi2016generalized}, \citet{herskovic2016common}
and \citet{renault2016apt}. Notice that the volatility co-movement does not arise from omitted factors in the AFM but from assuming a genuine\
factor structure in the idiosyncratic volatility processes.} For example, this technique would be valid for our illustrative example above. Yet, while the
QFA approach is able to recover the whole QFM structure for more general DGPs than the previous model (see subsection 2.2), PCA-SQ fails to do so. It will also fail when the idiosyncratic errors do not have bounded eighth moments. Hence, to the best of our knowledge, our QFA approach becomes the first estimation procedure capable of dealing with these issues.

\bigskip \noindent\textbf{Structure of the Paper}

The rest of the paper is organized as follows. Section 2 defines QFM and provides a list of simple illustrative examples where the new QFM methodology applies. In Section 3, we present the QFA estimator and its computational algorithm, establish the average rates of convergence of the quantile-dependent factors and factor loadings, and propose two consistent selection criteria to choose the number of factors at each quantile. Section 4 introduces a kernel-smoothed version of the QFA estimators to derive their asymptotic distributions. Section 5 contains some Monte Carlo simulation results to evaluate the performance in finite samples of our estimation procedures relative to other alternative approaches with  different assumptions about the idiosyncratic error terms. Section 6 considers three empirical applications using three large panel datasets, where we document the relevance of extra factors in causal analysis, forecasting and economic interpretation of common factors. Finally, Section 7 concludes and suggests several avenues for further research. Proofs of the main results are collected in the Online Appendix.



\bigskip \noindent\textbf{Notations}

The Frobenius norm is denoted as $\Vert \cdot \Vert $. For a matrix A with
real eigenvalues, $\rho _{j}(A)$ denotes the $j$th largest eigenvalue. Following \cite{vanweak}, the symbol $\lesssim $ means \textquotedblleft left side bounded by a positive constant times the right side\textquotedblright\ (the symbol $\gtrsim $ is defined similarly), and $%
D(\cdot ,g,\mathcal{G})$ denotes the packing number of space $\mathcal{G}$ endowed with semimetric $g$.

\section{The Model and Some Illustrative Examples}

This section starts by introducing the main definitions to be used throughout the paper. Next, we show how to derive the QFM representation of several illustrative DGPs exhibiting different factor structures.

\subsection{Quantile Factor Models}

\label{sectionmodel}

Suppose that the observed variable $X_{it},$ with $i=1,2,..,N$ and $%
t=1,2,...,T$, has the following QFM structure at some $\tau \in (0,1)$:
\begin{equation*}
\mathsf{Q}_{X_{it}}[\tau |f_t(\tau)]=\lambda _{i}^{\prime }(\tau)f_{t}(\tau ),
\end{equation*}%
where the common factors $f_{t}(\tau )$ is a $r(\tau )\times 1$ vector of
unobservable random variables, $\lambda _{i}(\tau )$ is a $r(\tau )\times 1$
vector of non-random factor loadings with $r(\tau)\ll N$. Note that in the QFM defined above, the factors, the loadings, and the number of factors are all allowed to be quantile-dependent.

Alternatively, the above equation implies that
\begin{equation}
X_{it}=\lambda _{i}^{\prime }(\tau )f_{t}(\tau )+u_{it}(\tau ),  \label{model}
\end{equation}%
where the quantile-dependent idiosyncratic error $u_{it}(\tau )$
is assumed to satisfy the following quantile restrictions:
\begin{equation*}
P[u_{it}(\tau )\leq 0|f_t(\tau)]=\tau .
\end{equation*}%

\subsection{Examples}

In this section we provide a few illustrative examples of how QFMs can be derived from different specifications of location-scale shift models and related ones. The goal of these simple illustrations is to show instances where the standard AFM methodology may fail to capture the full factor structure, therefore requiring the use of the alternative QFM approach.
\begin{exm}
\textbf{Location-shift model.} $X_{it}=\alpha _{i}f_{1t}+\epsilon _{it}$,
where $\{\epsilon _{it}\}$ are zero-mean i.i.d errors independent of $\{f_{1t}\}$
with cumulative distribution function (CDF) $\mathsf{F}_{\epsilon }$. Let $\mathsf{Q}_{\epsilon
}(\tau )=\mathsf{F}_{\epsilon }^{-1}(\tau )=\inf \{c:\mathsf{F}_{\epsilon
}(c)\leq \tau \}$ be the quantile function of $\epsilon_{it}$. Moreover, assume that the median of $\epsilon_{it}$ is 0, i.e., $\mathsf{Q}_{\epsilon }(0.5 )=0$, 
then this simple model has a QFM representation \eqref{model} by defining $\lambda _{i}(\tau )=[\mathsf{Q}_{\epsilon }(\tau ),\alpha _{i}]^{\prime }$, $f_{t}(\tau )=[1,f_{1t}]^{\prime }$ for $\tau\neq 0.5$, and $\lambda _{i}(\tau )=\alpha _{i}$, $f_{t}(\tau )=f_{1t}$ for $\tau= 0.5$. However, note that the standard estimation method (PCA) for
this AFM may not be consistent if the distribution of $\epsilon _{it}$
has heavy tails. For example, Assumption C of \textcolor{blue}{BN (2002)} requires $%
\mathbb{E}[\epsilon _{it}^{8}]<\infty $, which is not satisfied if, e.g. $%
\epsilon _{it}$ follows the standard Cauchy or some Pareto distributions.
\end{exm}

\begin{exm}
\textbf{Location-scale shift model (same sign-restricted factor). }$%
X_{it}=\alpha _{i}f_{1t}+\eta_{i}f_{1t}\epsilon _{it}$, where $\eta_i f_{1t}>0$ for all $i,t$ and $\{\epsilon _{it}\}$ are defined as in Example 1. This model has a QFM
representation \eqref{model} by defining $\lambda _{i}(\tau )=\eta_{i}\mathsf{Q}%
_{\epsilon }(\tau )+\alpha _{i}$ and $f_{t}(\tau )=f_{1t}$ for all $\tau$, such that the
loadings of the factor $f_{1t}$ are the only quantile-dependent objects.
\end{exm}

\begin{exm}
\textbf{Location-scale \textit{shift} model (different factors). }$%
X_{it}=\alpha _{i}^{\prime }f_{1t}+(\eta _{i}^{\prime
}f_{2t})\epsilon _{it}$, where $\{\epsilon _{it}\}$ are defined as in
Example 1, $\alpha _{i},f_{1t}\in \mathbb{R}^{r_{1}}$, $\eta _{i},$ $%
f_{2t}\in \mathbb{R}^{r_{2}}$, and $\eta _{i}^{\prime }f_{2t}>0$. When $%
f_{1t}$ and $f_{2t}$ do not share common elements, this model has a
QFM representation \eqref{model} with $\lambda _{i}(\tau )=[\alpha
_{i}^{\prime },\eta _{i}^{\prime }\mathsf{Q}_{\epsilon }(\tau )]^{\prime }$%
, $f_{t}(\tau )=[f_{1t}^{\prime },f_{2t}^{\prime }]$ for $\tau\neq 0.5$,  and $\lambda _{i}(\tau )=\alpha
_{i}$, $f_{t}(\tau )=f_{1t}$ for $\tau= 0.5$. 
\end{exm}

\begin{exm}
\textbf{Location-scale shift model with an idiosyncratic error and its cube.}
$X_{it}=\alpha _{i}f_{1t}+f_{2t}\epsilon _{it}+c_{i}f_{3t}\epsilon _{it}^{3}$%
, where $\epsilon _{it}$ is a standard normal random variable whose CDF is denoted as $\Phi(\cdot)$. Let $%
f_{2t},f_{3t},c_{i}$ be positive, then $X_{it}$ has an equivalent
representation in form of \eqref{model} with $\lambda _{i}(\tau )=[\alpha
_{i},\Phi ^{-1}(\tau ),c_{i}\Phi ^{-1}(\tau )^{3}]^{\prime }$, $f_{t}(\tau )=(f_{1t},f_{2t},f_{3t})'$ for $\tau\neq 0.5$, and $\lambda _{i}(\tau )=\alpha_{i}$, $f_t(\tau)=f_{1t}$ for $\tau=0.5$. In particular, if $c_{i}=1$
for all $i$ and noticing that the mapping $\tau \mapsto $ $\Phi ^{-1}(\tau
)^{3}$ is strictly increasing, then we have for $\tau\neq 0.5$, $Q_{X_{it}}[\tau
|f_t(\tau)]=\alpha _{i}f_{1t}+\Phi ^{-1}(\tau )\cdot \lbrack
f_{2t}+f_{3t}\Phi ^{-1}(\tau )^{2}],$ so that there exists a QFM
representation \eqref{model} with $\lambda _{i}(\tau )=[\alpha _{i},\Phi
^{-1}(\tau )]^{\prime }$ and $f_{t}(\tau )=[f_{1t},f_{2t}+f_{3t}\Phi
^{-1}(\tau )^{2}]^{\prime }$ for $\tau\neq 0.5$. Notice that in this case, the second factor in
$f_{t}(\tau )$, $f_{2t}+f_{3t}\Phi ^{-1}(\tau )^{2}$, is quantile dependent even for $\tau\neq 0.5$.
\end{exm}
Not surprisingly, the standard AFM methodology based on PCA only works in Example 1, insofar as the idiosyncratic errors satisfy certain moment conditions. In the remaining examples, PCA will only yield consistent estimates of those factors shifting the locations; however (except in Example 2), it will fail to capture those extra factors which shift quantiles other than the means, or their corresponding quantile-varying loadings. In the sequel, QFA is therefore proposed as a new estimation procedure to estimate both sets of quantile-dependent objects in QFM.

\section{Estimators and their Asymptotic Properties}
To simplify the notations, we suppress hereafter the dependence of $%
f_{t}(\tau ),\lambda _{i}(\tau ),r(\tau )$ and $u_{it}(\tau )$ on $\tau $,
so that the QFM in \eqref{model} is rewritten as:
\begin{equation}
X_{it}=\lambda _{i}^{\prime }f_{t}+u_{it},\quad P[u_{it}\leq 0|f_{t}]=\tau ,
\label{model2}
\end{equation}%
where $\lambda _{i},f_{t}\in \mathbb{R}^{r}$. Suppose that we have a sample
of observations $\{X_{it}\}$ generated by \eqref{model2} for $i=1,\ldots ,N,$
and $t=1,\ldots ,T$, where the realized values of $\{f_{t}\}$ are $%
\{f_{0t}\} $ and the true values of $\{\lambda _{i}\}$ are $\{\lambda _{0i}\}
$. We take a fixed-effects approach by treating $\{\lambda _{0i}\}$ and $%
\{f_{0t}\}$ as parameters to be estimated, and our asymptotic analysis is conditional on $\{f_{0t}\}$. In Section 3.1, we consider the
estimation of $\{\lambda _{0i}\}$ and $\{f_{0t}\}$ while $r$ is assumed to
be known. Finally, Section 3.2 deals with the estimation of $r$ for each
quantile.

\subsection{Estimating Factors and Loadings}

It is well known in the literature on factor models that $\{\lambda _{0i}\}$
and $\{f_{0t}\}$ cannot be separately identified without imposing
normalizations (see \textcolor{blue}{BN 2002}). Without loss of
generality, we choose the following normalizations:
\begin{equation}  \label{normalization}
\frac{1}{T}\sum_{t=1}^{T}f_{t}f_{t}^{\prime }=\mathbb{I}_{r},\quad \frac{1}{N%
}\sum_{i=1}^{N}\lambda _{i}\lambda _{i}^{\prime }\text{ is diagonal with
non-increasing diagonal elements.}
\end{equation}

Let $M=(N+T)r$, $\theta =(\lambda _{1}^{\prime },\ldots ,\lambda
_{N}^{\prime },f_{1}^{\prime },\ldots ,f_{T}^{\prime })^{\prime }$, and $%
\theta _{0}=(\lambda _{01}^{\prime },\ldots ,\lambda _{0N}^{\prime
},f_{01}^{\prime },\ldots ,f_{0T}^{\prime })^{\prime }$ denotes the vector
of true parameters, where we also suppress the dependence of $\theta $ and $%
\theta _{0}$ on $M$ to save notation. Let $\mathcal{A},\mathcal{F}%
\subset \mathbb{R}^{r}$ and define:
\begin{equation*}
\Theta ^{r}=\left\{ \theta \in \mathbb{R}^{M}:\lambda _{i}\in \mathcal{A}%
,f_{t}\in \mathcal{F}\text{ for all }i,t,\{\lambda _{i}\}\text{ and }%
\{f_{t}\}\text{ satisfy the normalizations in \eqref{normalization}}\right\}
.
\end{equation*}%
Further, define:
\begin{equation*}
\mathbb{M}_{NT}(\theta )=\frac{1}{NT}\sum_{i=1}^{N}\sum_{t=1}^{T}\rho _{\tau
}(X_{it}-\lambda _{i}^{\prime }f_{t}),
\end{equation*}%
where $\rho _{\tau }(u)=(\tau -\mathbf{1}\{u\leq 0\})u$ is the check
function. The QFA estimator of $\theta _{0}$ is defined as:
\begin{equation*}
\hat{\theta}=(\hat{\lambda}_{1}^{\prime },\ldots ,\hat{\lambda}_{N}^{\prime
},\hat{f}_{1}^{\prime },\ldots ,\hat{f}_{T}^{\prime })^{\prime }=\argmin%
_{\theta \in \Theta ^{r}}\mathbb{M}_{NT}(\theta ).
\end{equation*}%
It is obvious that the way in which our estimator is related to the PCA estimator studied by \textcolor{blue}{BN (2002)} and %
\citet{bai2003inferential} is analogous to how QR is related to standard least-squares regressions. However, unlike %
\citet{bai2003inferential}'s PCA estimator, our estimator $\hat{\theta}$ does not yield an analytical closed form. This makes it difficult not only to find a computational algorithm that would yield the estimator, but also the
analysis of its asymptotic properties. In the sequel, we introduce a computational algorithm called \textit{iterative quantile regression} (IQR, hereafter) that can effectively find the stationary points of the object function. In parallel, Theorem 1 shows that $\hat{\theta}$ achieves the same convergence rate as the PCA estimators for AFM.

To describe the algorithm, let $\Lambda =(\lambda _{1},\ldots ,\lambda
_{N})^{\prime }$, $F=(f_{1},\ldots ,f_{T})^{\prime }$, and define the
following averages:
\begin{equation*}
\mathbb{M}_{i,T}(\lambda ,F)=\frac{1}{T}\sum_{t=1}^{T}\rho _{\tau
}(X_{it}-\lambda ^{\prime }f_{t})\quad \text{ and }\quad \mathbb{M}%
_{t,N}(\Lambda ,f)=\frac{1}{N}\sum_{i=1}^{N}\rho _{\tau }(X_{it}-\lambda
_{i}^{\prime }f).
\end{equation*}%
Note that we have $\mathbb{M}_{NT}(\theta )=N^{-1}\sum_{i=1}^{N}\mathbb{M}%
_{i,T}(\lambda _{i},F)=T^{-1}\sum_{t=1}^{T}\mathbb{M}_{t,N}(\Lambda ,f_{t})$%
. The main difficulty in finding the global minimum of $\mathbb{M}_{NT}$ is
that this object function is not convex in $\theta $. However, for given $F$%
, $\mathbb{M}_{i,T}(\lambda ,F)$ happens to be convex in $\lambda $ for each
$i$ and likewise, for given $\Lambda $, $\mathbb{M}_{t,N}(\Lambda ,f)$ is
convex in $f$ for each $t$. Thus, both optimization problems can be efficiently solved by various linear programming methods (see Chapter 6 of  \citealt{koenker2005quantile}). Based on this observation, we propose the following iterative procedure:

\noindent {\textbf{Iterative quantile regression (IQR):} }\newline
\noindent {Step 1: }Choose random starting parameters: $F^{(0)}$. \newline
\noindent {Step 2: }Given $F^{(l-1)}$, solve $\lambda _{i}^{(l-1)}=\argmin%
_{\lambda }\mathbb{M}_{i,T}(\lambda ,F^{(l-1)})$ for $i=1,\ldots ,N$; given $%
\Lambda ^{(l-1)}$, solve $f_{t}^{(l)}=\argmin_{f}\mathbb{M}_{t,N}(\Lambda
^{(l-1)},f)$ for $t=1,\ldots ,T$. \newline
\noindent {Step 3: }For $l=1,\ldots ,L$, iterate the second step until $%
\mathbb{M}_{NT}(\theta ^{(L)})$ is close to $\mathbb{M}_{NT}(\theta
^{(L-1)}) $, where $\theta ^{(l)}=(\text{vech}(\Lambda ^{(l)})^{\prime },%
\text{vech}(F^{(l)})^{\prime })^{\prime }$. \newline
\noindent {Step 4: }Normalize $\Lambda ^{(L)}$ and $F^{(L)}$ so that they satisfy the normalizations in \eqref{normalization}.

To see the connection between the IQR algorithm and the PCA estimator of \citet{bai2003inferential}, suppose that $r=1$, and
replace the check function in the IQR algorithm by the least-squares loss function. Then, it is
easy to show that the second step of the algorithm above yields $\Lambda
^{(l-1)}=(X'F^{ (l-1)})/\Vert F^{(l-1)}\Vert ^{2}$ and $%
F^{(l)}=(X\Lambda ^{(l-1)})/\Vert \Lambda ^{(l-1)}\Vert ^{2}=XX'F^{(l-1)}/C_{l-1}$, where $X$ is the $T\times N$ matrix with elements $%
\{X_{it}\}$, and $C_{l}=\Vert F^{(l)}\Vert ^{2}\cdot \Vert \Lambda
^{(l)}\Vert ^{2}$. Thus, with proper normalizations at each step, the iterative procedure is equivalent to the
well-known \textit{power method} of \citet{hotelling1933analysis}, and the sequence $F^{(0)},F^{(1)},\ldots $ will converge to the
eigenvector associated with the largest eigenvalue of $XX^{\prime }$. In the more general case $r>1$, if we replace the check function in the IQR algorithm by the least-squares loss function and normalize $F^{(l-1)},\Lambda^{(l-1)}$ to satisfy \eqref{normalization} at step 2, it can be shown that the above iterative procedure is similar to the method of \textit{orthogonal iteration} (see Section 7.3.2 of \citealt{golub2013matrix}) for calculating the eigenvectors associated with the $r$ largest eigenvalues of $XX'$, which is the PCA estimator of \citet{bai2003inferential}. Therefore, the IQR algorithm and its corresponding QFA estimator can be viewed as an extension of PCA to QFM.

Similar algorithms have been proposed in the machine learning literature to reduce the dimensions for binary data, where the check function is replaced by some smooth nonlinear link functions, e.g. %
\citet{collins2002generalization}. However, unlike PCA, whether such methods guarantee finding the global minimum remains an important open question. Nonetheless, in all of our Monte Carlo simulations we found that the QFA estimators of the factors using the IQR algorithm always converge to the space of the true factors, which is somewhat reassuring in this respect.

To prove the consistency of the QFA estimator $\hat{\theta}$, we make the following assumptions:

\begin{ass}
\noindent (i) $\mathcal{A}$ and $\mathcal{F}$ are compact sets and $\theta
_{0}\in \Theta ^{r}$. In particular, $N^{-1}\sum_{i=1}^{N}\lambda
_{0i}\lambda _{0i}^{\prime }=\text{diag}(\sigma _{N1},\ldots ,\sigma _{Nr})$
with $\sigma _{N1}\geq \sigma _{N2}\cdots \geq \sigma _{Nr}$, and $\sigma
_{Nj}\rightarrow \sigma _{j}$ as $N\rightarrow \infty $ for $j=1,\ldots ,r$
with $\infty >\sigma _{1}>\sigma _{2}\cdots >\sigma _{r}>0$.\newline
(ii) The conditional density function of $u_{it}$ given $%
\{f_{0t}\}$, denoted as $\mathsf{f}_{it}$, is continuous, and satisfies that: for any compact set $C\subset \mathbb{R}$ and any $u\in C$, there exists a positive constant $\underline{\mathsf{f}}>0$ (depending on $C$) such that $\mathsf{f}_{it}(u)\geq
\underline{\mathsf{f}}$ for all $i,t$. \newline
(iii) Given $\{f_{0t},1\leq t\leq T\}$, $\{u_{it},1\leq i \leq N, 1\leq t\leq T\}$ is independent across $i$ and $t$.
\end{ass}
Assumptions 1 (i) is essentially the \textit{strong factors} assumption that is standard in the literature (see Assumption B of \citealt{bai2003inferential}). The requirement that $\sigma_1,\ldots,\sigma_r$ are distinct is similar to Assumption G of \citet{bai2003inferential}, which is a convenient assumption to order the factors. Assumptions 1 (ii) and (iii) are similar to (C1) and (C2) in \textcolor{blue}{AB (2020)}, except that we do not require moments of $u_{it}$ to exist. Also notice that Assumption (iii), which allows for both cross-sectional and time series heteroskedasticity, requires the idiosyncratic errors to be mutually independent. This stems from the use of Hoeffding's inequality in the proofs of some results, which provides a sub-Gaussian tail bound for the sum of bounded independent random variables. There have been attempts to relax this assumption (see Remark 1.4 below) but it is difficult to characterize the minimal set of conditions that the error terms should satisfy to achieve the sub-Gaussian inequality required in our proofs. Notice, however, that in exchange for the independence assumption, we can dispense with the bounded moment conditions in the idiosyncratic terms, whose violation would render PCA invalid. At any rate, in sub-section 5.2 we run some Monte Carlo simulation on the performance of our QFA estimation when error terms are allowed to exhibit mild cross-sectional and serial dependence in order to check the robustness of our results to these features. 

%
%
%

Write $\hat{\Lambda}=(\hat{\lambda}_{1},\ldots ,\hat{\lambda}_{N})^{\prime }$%
, $\Lambda _{0}=(\lambda _{01},\ldots ,\lambda _{0N})^{\prime }$, $\hat{F}=(%
\hat{f}_{1},\ldots ,\hat{f}_{T})^{\prime }$, $F_{0}=(f_{01},\ldots
,f_{0T})^{\prime }$, and let $L_{NT}=\min \{\sqrt{N},\sqrt{T}\}$. The
following theorem provides the average rate of convergence of $\hat{\Lambda}$
and $\hat{F}$.

\begin{thm}
Under Assumption 1, there exists a diagonal matrix $\Sz\in \mathbb{R}^{r\times r}$ whose diagonal elements are either $1$ or $-1$, such that as $N,T\rightarrow\infty$,
\begin{equation*}
\| \hat{\Lambda} - \Lambda_0 \Sz \|/\sqrt{N} = O_P(1/L_{NT}) \quad \text{ and }
\quad \| \hat{F}-F_0 \Sz\|/\sqrt{T} = O_P(1/L_{NT}) .
\end{equation*}
\end{thm}
The sign matrix $\Sz$ appears in the above result due the intrinsic sign indeterminacy of factors and loadings -- that is, the factor structure remains  unchanged if a factor and its loading are both multiplied by $-1$ (e.g., see Theorem 1.b of \citealt{stock2002forecasting} for a similar result). 

\noindent \textbf{Remark 1.1:} Since our proof strategy is substantially
different from that of \textcolor{blue}{BN (2002)}, we briefly sketch the
main ideas underlying our proof here. To facilitate the discussion, for any $\theta
_{a},\theta _{b}\in \Theta ^{r}$ define the semimetric $d$ by:
\begin{equation*}
d(\theta _{a},\theta _{b})=\sqrt{\frac{1}{NT}\sum_{i=1}^{N}\sum_{t=1}^{T}(%
\lambda _{ai}^{\prime }f_{at}-\lambda _{bi}^{\prime }f_{bt})^{2}}=\frac{1}{%
\sqrt{NT}}\left\Vert \Lambda _{a}F_{a}^{\prime }-\Lambda _{b}F_{b}^{\prime
}\right\Vert ,
\end{equation*}%
and let
\begin{equation*}
\bar{\mathbb{M}}_{NT}(\theta )=\frac{1}{NT}\sum_{i=1}^{N}\sum_{t=1}^{T}%
\mathbb{E}[\rho _{\tau }(X_{it}-\lambda _{i}^{\prime }f_{t})].
\end{equation*}%
The semimetric $d$ plays an important role in our asymptotic analysis. We
first show that $d(\hat{\theta},\theta _{0})=o_{P}(1)$. Next, it can be shown that:
\begin{equation}
\bar{\mathbb{M}}_{NT}(\hat{\theta})-\bar{\mathbb{M}}_{NT}(\theta
_{0})\gtrsim d^{2}(\hat{\theta},\theta _{0}),  \label{rmk1_1_1}
\end{equation}%
and that for sufficiently small $\delta >0$,
\begin{equation}
\mathbb{E}\left[ \sup_{\theta \in \Theta ^{r}(\delta )}\left\vert \mathbb{M}%
_{NT}(\theta )-\bar{\mathbb{M}}_{NT}(\theta )-\mathbb{M}_{NT}(\theta _{0})+%
\bar{\mathbb{M}}_{NT}(\theta _{0})\right\vert \right] \lesssim \frac{\delta
}{L_{NT}},  \label{rmk1_1_2}
\end{equation}%
where $\Theta ^{r}(\delta )=\{\theta \in \Theta ^{r}:d(\theta ,\theta
_{0})\leq \delta \}$. Intuitively, the above two inequalities and $d(\hat{%
\theta},\theta _{0})=o_{P}(1)$ imply that $d^{2}(\hat{\theta},\theta
_{0})\lesssim d(\hat{\theta},\theta _{0})/L_{NT}$, or $d(\hat{\theta},\theta
_{0})\lesssim L_{NT}^{-1}$. Then, the desired results follow from the fact
that $\Vert \hat{\Lambda}-\Lambda _{0}\Sz\Vert /\sqrt{N}+\Vert \hat{F}%
-F_{0}\Sz\Vert /\sqrt{T}\lesssim d(\hat{\theta},\theta _{0})$.

Inequality \eqref{rmk1_1_1} follows easily from a Taylor expansion of $\bar{%
\mathbb{M}}_{NT}(\hat{\theta})$ around $\theta _{0}$, together with Assumption 1(ii).
It is worth stressing that the proof of \eqref{rmk1_1_2} requires the
chaining argument which is commonly used in the theory of empirical processes. In particular, using Hoeffding's inequality and the fact that $%
|\rho _{\tau }(u)-\rho _{\tau }(v)|\leq 2|u-v|$, it can be shown that, for any given $\theta
_{a},\theta _{b}\in \Theta ^{r}$,
\begin{equation}
P\left[ \sqrt{NT}\left\vert \mathbb{M}_{NT}(\theta_a )-\bar{\mathbb{M}}%
_{NT}(\theta_a )-\mathbb{M}_{NT}(\theta _{b})+\bar{\mathbb{M}}_{NT}(\theta
_{b})\right\vert \geq c\right] \leq e^{-\frac{c^{2}}{Kd^{2}(\theta_a ,\theta
_{b})}}  \label{rmk1_1_3}
\end{equation}%
for some constant $K$. Then, along the lines of Theorem 2.2.4 of %
\citet{vanweak}, it follows that the left-hand side of \eqref{rmk1_1_2} is
bounded (up to a positive constant) by $\int_{0}^{\delta }\sqrt{\log D(\epsilon ,d,\Theta ^{r}(\delta ))}%
d\epsilon /\sqrt{NT}$. Finally, we can prove that $\int_{0}^{\delta }\sqrt{%
\log D(\epsilon ,d,\Theta ^{r}(\delta ))}d\epsilon \lesssim \delta \sqrt{M}$, from
which inequality \eqref{rmk1_1_2} follows.


\vspace{0.3cm} \noindent \textbf{Remark 1.2:} Compared to %
\textcolor{blue}{BN (2002)}, recall that, in exchange for Assumption 1(iii) we do not require any moment of $%
u_{it}$ to be finite. Thus, for the canonical AFM (e.g., Example 1) where the idiosyncratic errors have median equal to zero and satisfy Assumption 1(iii), our estimator for the case $\tau =0.5$ can be interpreted as a least absolute deviation (LAD) estimator which is robust to heavy tails and outliers. In relation to this issue, it is important to point out that the LAD estimator is related to robust PCA in the machine learning literature that aims to recover a low rank matrix from a large panel of observables. For example, the \textit{Principal Components Pursuit} method proposed by \citet{candes2011robust} features a combination of the $L_1$ norm (as in LAD) and a nuclear norm on the low rank matrix (see Chapter 3 of \citealt{vidal2016generalized} and \citealt{bai2019rank} for other robust PCA methods). In section 5 below, we will illustrate the robustness of the LAD estimator relative to the PCA estimator by Monte Carlo simulations. 

\vspace{0.3cm} \noindent \textbf{Remark 1.3:} If the true parameters do not
satisfy the normalizations \eqref{normalization}, they can still be in the
space $\Theta ^{r}$ after some normalizations. Let $H_{NT}$ be a $r\times r$
invertible matrix and define $\bar{f}_{0t}=H_{NT}^{\prime }f_{0t}$, $\bar{%
\lambda}_{0i}=(H_{NT})^{-1}\lambda _{0i}$. Note that $\lambda _{0i}^{\prime
}f_{0t}=\bar{\lambda}_{0i}^{\prime }\bar{f}_{0t}$. For $\{\bar{f}_{0t}\}$
and $\{\bar{\lambda}_{0i}\}$ to satisfy the normalizations %
\eqref{normalization}, we require:
\begin{equation*}
\frac{1}{T}\sum_{t=1}^{T}\bar{f}_{0t}\bar{f}_{0t}^{\prime }=H_{NT}^{\prime
}\Sigma _{T,F}H_{NT}=\mathbb{I}_{r}\quad \text{and}\quad \frac{1}{N}%
\sum_{i=1}^{N}\bar{\lambda}_{0i}\bar{\lambda}_{0i}^{\prime
}=(H_{NT})^{-1}\Sigma _{N,\Lambda }(H_{NT}^{\prime })^{-1}=\mathbb{D}_{N},
\end{equation*}%
where $\Sigma _{T,F}=T^{-1}\sum_{t=1}^{T}f_{0t}f_{0t}^{\prime }$, $\Sigma
_{N,\Lambda }=\frac{1}{N}\sum_{i=1}^{N}\lambda _{0i}\lambda _{0i}^{\prime }$%
, and $\mathbb{D}_{N}$ is a diagonal matrix with non-increasing diagonal
elements. The above equalities imply that:
\begin{equation*}
\Sigma _{T,F}^{1/2}\Sigma _{N,\Lambda }\Sigma _{T,F}^{1/2}\cdot \Sigma
_{T,F}^{1/2}H_{NT}=\Sigma _{T,F}^{1/2}H_{NT}\cdot \mathbb{D}_{N}.
\end{equation*}%
Thus, the rotation matrix $H_{NT}$ can be chosen as $\Sigma
_{T,F}^{-1/2}\Gamma _{NT}$, where $\Gamma _{NT}$ is the matrix of
eigenvectors of $\Sigma _{T,F}^{1/2}\Sigma _{N,\Lambda }\Sigma _{T,F}^{1/2}$. Note that when the eigenvalues of $\Sigma _{T,F}^{1/2}\Sigma _{N,\Lambda }\Sigma _{T,F}^{1/2}$ are distinct, its eigenvectors are unique up to signs, i.e., each eigenvector can be replaced by the negative of itself. As a result, Theorem 1 can be stated as follows:
\begin{equation*}
\Vert \hat{\Lambda}-\Lambda _{0}(H_{NT}^{\prime })^{-1}\Sz\Vert /\sqrt{N}%
=O_{P}(1/L_{NT})\quad \text{and}\quad \Vert \hat{F}-F_{0}H_{NT}\Sz\Vert /\sqrt{T%
}=O_{P}(1/L_{NT}),
\end{equation*}
where $\Sz$ is the diagonal matrix defined above. Notice that the rotation matrix $H_{NT}$ is slightly different from the rotation matrix of \citet{bai2003inferential}. Moreover, because both $\lambda_{0i}$ and $f_{0t}$ are $\tau$-dependent, $H_{NT}$ also varies across quantiles, although we did not make it explicitly quantile dependent in the previous discussion to simplify notation.

\vspace{0.3cm} \noindent \textbf{Remark 1.4:} Compared to %
\textcolor{blue}{BN (2002)}, our Assumption 1(iii) is admittedly strong.
However, note that this assumption is made conditional on $\{f_{0t}\}$, so
cross-sectional and temporal dependence of $u_{it}$ due to the common factors are still
allowed for. Moreover, the independence assumption is only used to establish
the sub-Gaussian inequality \eqref{rmk1_1_3}. Thus, Assumption 1(iii) can be
relaxed as long as the sub-Gaussian inequality holds.\footnote{%
See \citet{van2002hoeffding} for the properties of Hoeffding inequalities for martingales.}

\bigskip

\subsection{Selecting the Number of Factors}

In the previous section, we assumed the number of quantile-dependent factors $r(\tau )$ to be known at each $\tau $. In this subsection we propose two
different procedures to select the correct number of factors at each quantile with probability approaching one. The first one selects the number of factors by rank minimization while the second one uses information criteria (IC). As before, the dependence of the quantile-dependent objects on $\tau $, including $r(\tau )$%
, is suppressed for notational simplicity.

\subsubsection{Model Selection by Rank Minimization}

Let $k$ be a positive integer larger than $r$, and $\mathcal{A}^{k}$ and $%
\mathcal{F}^{k}$ be compact subsets of $\mathbb{R}^{k}$. In particular, let us
assume that $[ \lambda_{0i}^{\prime }\quad \mathbf{0}_{1\times (k -r ) }
]^{\prime }\in \mathcal{A}^{k} $ for all $i$.

Let $\lambda _{i}^{k},f_{t}^{k}\in \mathbb{R}^{k}$ for all $i,t$ and write $%
\theta ^{k}=(\lambda _{1}^{k^{\prime }},\ldots ,\lambda _{N}^{k^{\prime
}},f_{1}^{k^{\prime }},\ldots ,f_{T}^{k^{\prime }})^{\prime }$, $\Lambda
^{k}=(\lambda _{1}^{k},\ldots ,\lambda _{N}^{k})^{\prime }$, $%
F^{k}=(f_{1}^{k},\ldots ,f_{T}^{k})^{\prime }$. Consider the following
normalizations:
\begin{equation}
\frac{1}{T}\sum_{t=1}^{T}f_{t}^{k}f_{t}^{k^{\prime }}=\mathbb{I}_{k},\quad
\frac{1}{N}\sum_{i=1}^{N}\lambda _{i}^{k}\lambda _{i}^{k^{\prime }}\text{ is
diagonal with non-increasing diagonal elements.}  \label{IC_norm}
\end{equation}%
Define $\Theta ^{k}=\{\theta ^{k}:\lambda _{i}^{k}\in \mathcal{A}%
^{k},f_{t}^{k}\in \mathcal{F}^{k},\text{ and }\lambda _{i}^{k},f_{t}^{k}%
\text{ satisfy \eqref{IC_norm}}\}$, and
\begin{equation*}
\hat{\theta}^{k}=(\hat{\lambda}_{1}^{k^{\prime }},\ldots ,\hat{\lambda}%
_{N}^{k^{\prime }},\hat{f}_{1}^{k^{\prime }},\ldots ,\hat{f}_{T}^{k^{\prime
}})^{\prime }=\argmin_{\theta ^{k}\in \Theta ^{k}}\frac{1}{NT}%
\sum_{i=1}^{N}\sum_{t=1}^{T}\rho _{\tau }(X_{it}-\lambda _{i}^{k^{\prime
}}f_{t}^{k}).
\end{equation*}%
Moreover, define $\hat{\Lambda}^{k}=(\hat{\lambda}_{1}^{k},\ldots ,\hat{%
\lambda}_{N}^{k})^{\prime }$ and write
\begin{equation*}
(\hat{\Lambda}^{k})^{\prime }\hat{\Lambda}^{k}/N=\diag\left( \hat{\sigma}%
_{N,1}^{k},\ldots ,\hat{\sigma}_{N,k}^{k}\right) .
\end{equation*}%
The first estimator of the number of factors $r$ is defined as:
\begin{equation*}
\hat{r}_{\text{rank}}=\sum_{j=1}^{k}\mathbf{1}\{\hat{\sigma}%
_{N,j}^{k}>P_{NT}\},
\end{equation*}%
where $P_{NT}$ is a sequence that goes to 0 as $N,T\rightarrow \infty $. In
other words, $\hat{r}_{\text{rank}}$ is equal to the number of diagonal
elements of $(\hat{\Lambda}^{k})^{\prime }\hat{\Lambda}^{k}/N$ that are
larger than the threshold $P_{NT}$. We call $\hat{r}_{\text{rank}}$ the
\textit{rank-minimization estimator} because, as discussed below in Remark 2.1, it can be
interpreted as a rank estimator of $(\hat{\Lambda}^{k})^{\prime }\hat{\Lambda%
}^{k}/N$.

It can then be shown that:

\begin{thm}
Under Assumption 1, $P[\hat{r}_{\text{rank}}=r]\rightarrow 1$ as $%
N,T\rightarrow \infty$ if $k>r$, $P_{NT}\rightarrow 0$ and $P_{NT}L_{NT}^{2}\rightarrow \infty $.
\end{thm}

\vspace{0.3cm} \noindent \textbf{Remark 2.1:} In the proof of Theorem 2, we
show that for $k>r$, it holds that (up to sign)
\begin{equation*}
\left\Vert \hat{F}^{k,r}-F_{0}\right\Vert /\sqrt{T}=O_{P}(1/L_{NT})\quad
\text{ and }\quad \left\Vert \hat{\Lambda}^{k}-\Lambda _{0}^{\ast
}\right\Vert /\sqrt{N}=O_{P}(1/L_{NT}),
\end{equation*}%
where $\hat{F}^{k,r}$ is the first $r$ columns of $\hat{F}^{k}$ and $\Lambda
_{0}^{\ast }=[\Lambda _{0},\mathbf{0}_{N\times (k-r)}]$. It then follows
from Assumption 1 that $\hat{\sigma}_{N,j}^{k}\overset{p}{\rightarrow }%
\sigma _{j}>0$ for $j=1,\ldots ,r$ and $\hat{\sigma}_{N,j}^{k}=N^{-1}%
\sum_{i=1}^{N}\left( \hat{\lambda}_{i,j}^{k}\right) ^{2}=O_{P}(1/L_{NT}^{2})$ for $j=r+1,\ldots,k$. Thus, the first $r$ diagonal components of $(\hat{\Lambda}^{k})^{\prime }%
\hat{\Lambda}^{k}/N$ converge in probability to positive constants while the
remaining diagonal components are all $O_{P}(1/L_{NT}^{2})$. In other words,
$(\hat{\Lambda}^{k})^{\prime }\hat{\Lambda}^{k}/N$ converges to a matrix
with rank $r$, and $P_{NT}$ can be viewed as a cutoff value to choose the
asymptotic rank of $(\hat{\Lambda}^{k})^{\prime }\hat{\Lambda}^{k}/N$.

\subsubsection{Model Selection by Information Criteria}

The second estimator of $r$ is similar to the IC-based
estimator of \textcolor{blue}{BN (2002)}. Let $l$ denote a positive integer
smaller or equal to $k$, and $\mathcal{A}^{l}$ and $\mathcal{F}^{l}$ be
compact subsets of $\mathbb{R}^{l}$. In particular, for $l>r$, assume that $%
[ \lambda_{0i}^{\prime }\quad \mathbf{0}_{1\times (l -r ) } ]^{\prime }\in
\mathcal{A}^{l} $ for all $i$. Moreover, we can define $\Theta ^{l}, \hat{%
\theta}^{l} , \hat{f}^l_{t},\hat{\lambda}_i^l, \hat{F}^l$ and $\hat{\Lambda}%
^l$ in a similar fashion.

Define the IC-based estimator of $r$ as follows:
\begin{equation*}
\hat{r}_{\text{IC}} = \argmin_{1\leq l\leq k} \left[ \mathbb{M}_{NT} (\hat{%
\theta}^l ) +l\cdot P_{NT} \right].
\end{equation*}
We can show that:

\begin{thm}
Suppose Assumption 1 holds, and assume that for any compact set $C\subset \mathbb{R}$ and any $u\in C$, there exists $\bar{\mathsf{f}}>0$ (depending on $C$)
such that $
\mathsf{f}_{it}(u)\leq \bar{\mathsf{f}}$ for all $i,t$. Then $P[\hat{r}%
_{IC}=r]\rightarrow 1$ as $N,T\rightarrow \infty$ if $k>r$, $P_{NT}\rightarrow 0$ and $%
P_{NT}L_{NT}^{2}\rightarrow \infty $.
\end{thm}

\vspace{0.3cm} \noindent \textbf{Remark 3.1:} \textcolor{blue}{AB (2020)} obtain a similar result, but the difference with ours is that we only need the density function of the idiosyncratic errors to be uniformly bounded above and below, while \textcolor{blue}{AB (2020)} requires all the moments of the errors to be
bounded. The reason why we can obtain the same result here with less restrictions is that our proof is based on the innovative argument discussed
in Remark 1.1 and on the average convergence rate of the estimators, while the proof of \textcolor{blue}{AB (2020)} depends on the uniform convergence rate of the estimators.

\vspace{0.3cm} \noindent \textbf{Remark 3.2:} 
Let $X$ denote the $T\times N$ matrix of observed variables, and let $\check{F}%
^{l},\check{\Lambda}^{l}$ denote the matrices of PCA estimators of \textcolor{blue}{BN (2002)} when the number of factors is specified as $l$. Then \textcolor{blue}{BN (2002)}'s estimator of $r$ can be written as:
\begin{equation*}
\hat{r}=\argmin_{1\leq l\leq k}\hat{S}(l)\quad \text{ where }\quad \hat{S}%
(l)=(NT)^{-1}\left\Vert X-\check{F}^{l}\check{\Lambda}^{l^{\prime
}}\right\Vert ^{2}+l\cdot P_{NT},
\end{equation*}%
$k>r$, and $P_{NT}$ is defined as in Theorem 2 above. It can be shown that IC-based estimator $\hat{r}$ is equivalent to the number of diagonal elements in $\check{\Lambda}^{k^{\prime }}\check{%
\Lambda}^{k}/N$ that are larger than $P_{NT}$. Thus, the two seemly different estimators of the number of factors are equivalent in AFM. However, due to the differences of the object functions, such equivalence does not exist in QFM.

\vspace{0.3cm} \noindent \textbf{Remark 3.3:} The choice of $P_{NT}$ for $%
\hat{r}_{\text{rank}}$ and $\hat{r}_{\text{IC}}$ can be different in practice. In particular, it can differ from those penalties used by %
\textcolor{blue}{BN (2002)}. \textcolor{blue}{AB (2020)} choose
\begin{equation*}
P_{NT}=\log \left( \frac{NT}{N+T}\right) \cdot \frac{N+T}{NT}
\end{equation*}%
for $\hat{r}_{\text{IC}}$, similar to $IC_{p1}$ of \textcolor{blue}{BN (2002)}%
. However, as shown in \textcolor{blue}{AB's (2020)} simulation results, this choice does not
perform very well even for $N,T$ as large as 300.

\vspace{0.3cm} \noindent \textbf{Remark 3.4:} Even though $\hat{r}_{\text{%
rank}}$ and $\hat{r}_{\text{IC}}$ are both consistent estimators of $r$, the
computational cost of $\hat{r}_{\text{rank}}$ is much lower than that of $%
\hat{r}_{\text{IC}}$, because for $\hat{r}_{\text{rank}}$ we only estimate
the model once, while for $\hat{r}_{\text{IC}}$ we need to estimate the
model $k$ times. Thus, in the simulations and empirical applications we will focus on $\hat{r}_{\text{%
rank}}$, and we refer to \textcolor{blue}{AB (2020)} for the corresponding simulation results of $\hat{r}_{%
\text{IC}}$. In particular, we find that the choice
\begin{equation*}
P_{NT}=\hat{\sigma}^k_{N,1}\cdot \left( \frac{1}{L_{NT}^{2}}\right) ^{1/3}
\end{equation*}%
for $\hat{r}_{\text{rank}}$ works fairly well as long as $\min \{N,T\}$ is
100. This is also the value used in all of our simulations and applications.

\section{Estimators Based on Smoothed Quantile Regressions}

The derivation of the asymptotic distribution of the QFA estimator $\hat{\theta}$ becomes a difficult task due to the non-smoothness of the check function and the problem of
incidental parameters. As in the asymptotic analysis of conventional QR, one can expand the expected score function (which is smooth and continuously differentiable) and obtain a stochastic expansion for $\hat{%
\lambda}_{i}-\Sz\lambda _{0i}$; yet the following term appears in the
expansion:
\begin{equation}
\frac{1}{T}\sum_{t=1}^{T}\left\{ \left( \mathbf{1}\{X_{it}\leq \hat{\lambda}%
_{i}^{\prime }\hat{f}_{t}\}-\mathbb{E}[\mathbf{1}\{X_{it}\leq \hat{\lambda}%
_{i}^{\prime }\hat{f_{t}}\}]\right) \hat{f}_{t}-\left( \mathbf{1}%
\{X_{it}\leq \lambda _{0i}^{\prime }f_{0t}\}-\tau \right) f_{0t}\right\} .
\label{8.1}
\end{equation}%
The next step would be to show that the above expression is a higher-order term (i.e. $o_P(T^{-0.5})$) thus it does not affect the asymptotic distribution of $\hat{\lambda}_{i}$. However, due to the presence of the indicator functions in \eqref{8.1}, this is not an easy task. 
To see this, let's consider a similar problem for the PCA estimators of AFM. Let $%
\check{\lambda}_{i}$ and $\check{f}_{t}$ be the PCA estimators. In
the stochastic expansion of $\check{\lambda}_{i}-\lambda _{0i}$, the analogous term to \eqref{8.1} happens to be:
\begin{equation*}
\frac{1}{T}\sum_{t=1}^{T}\epsilon _{it}(\check{f}_{t}-f_{0t}),
\end{equation*}%
where $\epsilon _{it}$ is the idiosyncratic error in the AFM. Note that, based on the result $%
T^{-1}\sum_{t=1}^{T}\Vert \check{f}_{t}-f_{0t}\Vert =O_{P}(L_{NT}^{-1})$, one can only show that:
\[ \left\| \frac{1}{T}\sum_{t=1}^{T}\epsilon _{it}(\check{f}_{t}-f_{0t}) \right\| \leq \sqrt{ \frac{1}{T}\sum_{t=1}^{T}\epsilon_{it}^2 }\cdot \sqrt{ \frac{1}{T}\sum_{t=1}^{T}\| \check{f}_{t}-f_{0t} \|^2} =O_{P}(L_{NT}^{-1})  .\]
Instead, one has to use the stochastic expansion of $\check{f}_{t}- f_{0t}$ to show that $%
T^{-1}\sum_{t=1}^{T}\epsilon _{it}(\check{f}_{t}-f_{0t})=O_P(L_{NT}^{-2})$ (see
the proof of Lemma B.1 of \citealt{bai2003inferential}). Likewise, to show that \eqref{8.1} is $o_{P}(T^{-0.5})$, establishing the convergence rate of $\hat{f}_{t}-\Sz f_{0t}$ is not enough, and the stochastic expansion of $\hat{f}_{t}-\Sz f_{0t}$ is required. However, due the non-smoothness of the
indicator functions, it is not clear how to explore this stochastic expansion in \eqref{8.1}.

To overcome this problem, we proceed to define a new
estimator of $\theta _{0}$, denoted as $\tilde{\theta}$, relying on the
following smoothed quantile regressions (SQR):
\begin{equation*}
\tilde{\theta}=(\tilde{\lambda}_{1}^{\prime },\ldots ,\tilde{\lambda}%
_{N}^{\prime },\tilde{f}_{1}^{\prime },\ldots ,\tilde{f}_{T}^{\prime
})^{\prime }=\argmin_{\theta \in \Theta ^{r}}\mathbb{S}_{NT}(\theta ),
\end{equation*}%
where
\begin{equation*}
\mathbb{S}_{NT}(\theta )=\frac{1}{NT}\sum_{i=1}^{N}\sum_{t=1}^{T}\left[ \tau
-K\left( \frac{X_{it}-\lambda _{i}^{\prime }f_{t}}{h}\right) \right]
(X_{it}-\lambda _{i}^{\prime }f_{t}),
\end{equation*}
$K(z)=1-\int_{-1}^{z}k(z)dz$, $k(z)$ is a continuous function with support $%
[-1,1]$, and $h$ is a bandwidth parameter that goes to 0 as $N,T$ diverge.

Define
\begin{equation*}
\Phi_{i} =\lim_{T\rightarrow\infty} \frac{1}{T}\sum_{t=1}^{T} \mathsf{f}%
_{it}(0)f_{0t}f_{0t}^{\prime }\quad \text{ and } \quad\Psi_{t}=\lim_{N%
\rightarrow\infty} \frac{1}{N}\sum_{i=1}^{N} \mathsf{f}_{it}(0)\lambda_{0i}%
\lambda_{0i}^{\prime }
\end{equation*}
for all $i,t$. We impose the following assumptions:

\begin{ass}
\noindent Let $m\geq 8$ be a positive integer, \\
(i) $\Phi_{i}>0$ and $\Psi_{t}>0$ for all $i,t$. \newline
(ii) $\lambda_{0i}$ is an interior point of $\mathcal{A}$ and $f_{0t}$ is an
interior point of $\mathcal{F}$ for all $i,t$.\newline
(iii) $k(z)$ is symmetric around $0$ and twice continuously
differentiable. $\int_{-1}^{1} k(z)dz =1$, $\int_{-1}^{1}z^j
k(z)dz =0$ for $j=1,\ldots,m-1$ and $\int_{-1}^{1}z^m k(z)dz \neq 0$.\newline
(iv) $\mathsf{f}_{it}$ is $m+2$ times continuously differentiable. Let $%
\mathsf{f}_{it}^{(j)}(u) = (\partial /\partial u)^j\mathsf{f}_{it}(u) $ for $%
j=1,\ldots, m+2$. For any compact set $C\subset \mathbb{R}$ and any $u\in C$, there exists $-\infty<\underline{l}<\bar{l}<\infty$ such
that $\underline{l} \leq \mathsf{f}_{it}^{(j)}(u)\leq \bar{l} $ and $\underline{%
\mathsf{f}} \leq \mathsf{f}_{it}(u) \leq\bar{l} $ for $j=1,\ldots,m+2$ and
for all $i,t$. \newline
(v) As $N,T\rightarrow \infty$, $N\propto T$, $h\propto T^{-c}$ and $m^{-1}
< c < 1/6$. 
\end{ass}
The above conditions are standard in SQR, with the exception of $(v)$. Note that, like \citet{galvao2016smoothed}, we need $k(z)$ to be a higher-order kernel function to control the higher-order terms in the stochastic expansions of the estimators. However, \citet{galvao2016smoothed} assume that $m^{-1}<c<1/3$ (or $m\geq 4$), while we need $m^{-1}<c<1/6$ (or $m\geq 8$). The difference is due to the fact that the incidental parameters ($\lambda_{i}$ and $f_t$) in QFM enter the model interactively, while in the panel quantile models considered by these authors there are no interactive fixed-effects. 

Then, we can show that:
\begin{thm}
Under Assumptions 1 and 2, there exists a diagonal matrix $\Sz\in \mathbb{R}^{r\times r}$ whose diagonal elements are either $1$ or $-1$, such that
\begin{equation*}
\sqrt{T} ( \tilde{\lambda}_i -\Sz\lambda_{0i}) \overset{d}{\rightarrow}
\mathcal{N}(0, \tau(1-\tau)\Phi_i^{-2} )\quad \text{ and }\quad\sqrt{N} (%
\tilde{f}_t -\Sz f_{0t} ) \overset{d}{\rightarrow} \mathcal{N}(0,\tau(1-\tau)
\Psi_t^{-1} \Sigma_{\Lambda} \Psi_t^{-1})
\end{equation*}
for each $i$ and $t$, where $\Sigma_{\Lambda}=\diag(\sigma_1,\ldots,%
\sigma_r) $.
\end{thm}

\noindent \textbf{Remark 4.1:} Similar to the proof of Theorem 1, we can
show that
\begin{equation*}
\| \tilde{\Lambda} - \Lambda_0\Sz\|/\sqrt{N} = O_P(1/L_{NT})+O_P(h^{m/2}) \quad
\text{ and } \quad \| \tilde{F}-F_0\Sz\|/\sqrt{T} = O_P(1/L_{NT})+O_P(h^{m/2}),
\end{equation*}
where the extra $O_P(h^{m/2})$ term is due the approximation bias of the
smoothed check function. However, Assumption 2$(v)$ implies that $%
1/L_{NT}\gg h^{m/2}$, and then it follows that average convergence rates of $%
\tilde{\Lambda}$ and $\tilde{F}$ are both $L_{NT}$.

\noindent \textbf{Remark 4.2:} Similar to Theorems 1 and 2 of %
\citet{bai2003inferential}, we show that the new estimator is free of
incidental-parameter biases. That is, the asymptotic distribution of $\tilde{%
\lambda}_{i}$ is the same as if we would observe $\{f_{0t}\}$, and likewise
the asymptotic distribution of $\tilde{f}_{t}$ is the same as if $\{\lambda
_{0i}\}$ were observed. The proof of this result is not trivial. To see why
this is the case, first define $\varrho (u)=[\tau -K(u/h)]u$ and $\mathbb{S}%
_{i,T}(\lambda ,F)=T^{-1}\sum_{t=1}^{T}\varrho (X_{it}-\lambda ^{\prime
}f_{t})$, then we can write $\tilde{\lambda}_{i}=\argmin_{\lambda \in
\mathcal{A}}\mathbb{S}_{i,T}(\lambda ,\tilde{F})$. Expanding $\partial
\mathbb{S}_{i,T}(\tilde{\lambda}_{i},\tilde{F})/\partial \lambda $ around $%
(\lambda _{0i},F_{0})$ yields
\begin{multline}
\left( \frac{1}{T}\sum_{t=1}^{T}\varrho ^{(2)}(u_{it})f_{0t}f_{0t}^{\prime
}\right) (\tilde{\lambda}_{i}-\Sz\lambda _{0i})\approx \frac{1}{T}%
\sum_{t=1}^{T}\varrho ^{(1)}(u_{it})\Sz f_{0t}+\frac{1}{T}\sum_{t=1}^{T}\varrho
^{(1)}(u_{it})(\tilde{f}_{t}-\Sz f_{0t})  \label{eq8} \\
-\frac{1}{T}\sum_{t=1}^{T}\varrho ^{(2)}(u_{it})f_{0t}\lambda _{0i}^{\prime }(%
\tilde{f}_{t}-\Sz f_{0t}),
\end{multline}%
where $\varrho ^{(j)}(u)=(\partial /\partial u)^{j}\varrho (u)$. The key
step is to show that the last two terms on the right-hand side of the above
equation are both $o_{P}(1/\sqrt{T})$. This is relatively easier for the PCA
estimator of \citet{bai2003inferential}, since $(\tilde{f}_{t}-\Sz f_{0t})$ has an analytical
form (like e.g. in equation A.1 of \citealt{bai2003inferential}). In our case, we
would also need a stochastic expansion for $(\tilde{f}_{t}-\Sz f_{0t})$, which in turn depends on the stochastic expansion of $(%
\tilde{\lambda}_{i}-\Sz \lambda _{0i})$ due to the nature of factor models.
As in \citet{chen2018panel}, this problem can be partly solved by
showing that the expected Hessian matrix is asymptotically block-diagonal
(see Lemma 11 in the Online Appendix). %
However, the proof of \citet{chen2018panel} is only applicable to a special
infeasible normalization, namely $\sum_{i=1}^{N}\lambda _{0i}\lambda
_{i}=\sum_{t=1}^{T}f_{0t}f_{t}^{\prime }$, while our proof of Lemma 11
allows for normalization \eqref{normalization} and can be generalized to any
of the other normalizations considered by \citet{bai2013principal} that uniquely
pin down the rotation matrix.

\vspace{0.3cm} \noindent \textbf{Remark 4.3:} As discussed in Remark 1.3, if
the true parameters do not satisfy the normalizations \eqref{normalization},
the results of Theorem 3 can be stated as
\begin{equation*}
\sqrt{T} \left( \tilde{\lambda}_i - \Sz H_{NT}^{-1}\lambda_{0i}\right) \overset{d%
}{\rightarrow} \mathcal{N}\left(0, \tau(1-\tau)H^{-1}\Phi_i^{-1}
\Sigma_F\Phi_i^{-1} (H^{-1})'\right),
\end{equation*}
\begin{equation*}
\sqrt{N} \left( \tilde{f}_t -\Sz H_{NT}^{\prime }f_{0t} \right) \overset{d}{%
\rightarrow} \mathcal{N}\left(0,\tau(1-\tau) H^{\prime }\Psi_t^{-1}
\Sigma_{\Lambda} \Psi_t^{-1} H \right),
\end{equation*}
where $\Sz$ and $H_{NT}$ are defined in Remark 1.3, $\Sigma_F= \lim_{T\rightarrow\infty}\Sigma_{T,F}$, $\Sigma_{\Lambda}= \lim_{T\rightarrow\infty}\Sigma_{N,\Lambda}$, $H = \Sigma_F^{-1/2}\Gamma$,
and $\Gamma$ is the matrix of eigenvectors of $\Sigma_F^{1/2}\Sigma_\Lambda
\Sigma_F^{1/2}$.

\vspace{0.3cm}
\noindent\textbf{Remark 4.4: } Let $l(z)$ be a continuous kernel function with support $[-1,1]$ where $l^{(j)}(z) =\partial^j l(z)/\partial z^j$ exists and $|l^{(j)}(z)|$ is bounded for $j=1,2$. Let $b$ a bandwidth. Estimators for the asymptotic variance matrices of $\tilde{\lambda}_i$ and $\tilde{f}_t$ can be simply constructed as 
\[ \tilde{V}_{\lambda_i} =\tau(1-\tau)\tilde{\Phi}_i^{-2} \text{ where } \tilde{\Phi}_i = \frac{1}{Tb}\sum_{t=1}^{T} l(\tilde{u}_{it}/b) \cdot \tilde{f}_t\tilde{f}_t',\]
and
\[ \tilde{V}_{f_t}  =\tau(1-\tau)\tilde{\Psi}_t^{-1}\tilde{\Sigma}_{\Lambda}\tilde{\Psi}_t^{-1} \text{ where } \tilde{\Psi}_t = \frac{1}{Nb}\sum_{i=1}^{N} l(\tilde{u}_{it}/b) \cdot \tilde{\lambda}_i \tilde{\lambda}_i',  \quad \tilde{\Sigma}_{\Lambda} = \tilde{\Lambda}' \tilde{\Lambda}/N,\]
with $\tilde{u}_{it} = X_{it} -\tilde{\lambda}_i'\tilde{f}_t$. In Section A.5 of the Online Appendix we show that under Assumptions 1 and 2, the above estimators of the asymptotic covariance matrices are consistent if $b\rightarrow 0$ and $Nb^3\rightarrow\infty$. Note that this is different from the usual condition $Nb^2\rightarrow\infty$ in standard quantile regressions (see e.g. \citealt{powell1984least} and \citealt{angrist2006quantile}). Moreover, the above estimators are also consistent for the asymptotic covariance matrices discussed in Remark 4.3.

\vspace{0.3cm}
\noindent\textbf{Remark 4.5: }A restrictive DGP within class (1) would be a QFM where
the PCA factors coincide with the quantile factors and only the factor
loadings are quantile dependent. The representation for such restricted subset of QFM is as follows:
\begin{equation}
X_{it}=\lambda _{i}^{\prime }(\tau )f_{t}+u_{it}(\tau ),\text{ for }\tau \in
(0,1).
\end{equation}

As a result, the main objects of interest are the common factors and the
quantile-varying loadings. Notice that, if the factors $f_{t}$ were to be
observed, using standard QR of $X_{it}$ on $f_{t}$ would lead to consistent
and asymptotically normally distributed estimators of $\lambda _{i}(\tau )$
for each $i$ and $\tau \in (0,1)$. However, since $f_{t}$ are not
observable, a feasible two-stage approach is to first estimate the factors
by PCA, denoted as $\hat{f}_{PCA,t}$, and next run QR of $X_{it}$ on $\hat{f}_{PCA,t}$ to obtain
estimates of $\lambda _{i}(\tau )$ as follows:

\begin{equation}
\hat{\lambda}_{i}(\tau )=\argmin_{\lambda} T^{-1}%
\sum_{t=1}^{T}\rho _{\tau }(X_{it}-\lambda ^{\prime }\hat{f}_{PCA,t}).
\end{equation}

As explained in \citet{chen2017qfm}, unlike the QFA estimators (see Remark 1.2),
this two-stage procedure requires moments of the idiosyncratic term $u_{it}$
to be bounded in order to apply PCA in the first stage. However, an interesting result (see \citealt{chen2017qfm}, Theorem 2) is that the standard conditions on the relative asymptotics
of $N$ and $T$ allowing for the estimated factors to be treated as known do
not hold when applying this two-stage estimation approach. In effect, while
these conditions are $T^{1/2}/N\rightarrow 0$ for linear factor-augmented
regressions (see \citealt{bai2006confidence}) and $T^{5/8}/N\rightarrow 0$
for nonlinear factor-augmented regressions (\citealt{bai2008extremum}),
lack of smoothness in the object (check) function at the second stage
requires the stronger condition $T^{5/4}/N\rightarrow 0$. Moreover, Theorem
3 in \citet{chen2017qfm} shows how to run inference on the quantile-varying
loadings (e.g., testing the null that they are constant across all quantiles
or a subset of them).


\section{Finite Sample Simulations}

In this section we report the results from several Monte Carlo simulations regarding the performance of our proposed QFM methodology in finite samples. In particular, we focus on four relevant issues: (i) how well does our preferred estimator of the number of factors perform relative to other methods when the distribution of the idiosyncratic errors in an AFM exhibits heavy tails or outliers, (ii) how well do PCA and QFA estimate the true factors under the previous circumstances, (iii) how robust is the QFA estimation procedure when the errors terms are serially and cross-sectionally correlated, instead of being independent, and (iv) how good are the normal approximations given in Theorem 4 for the QFA estimators based on SQR. 

\subsection{Estimation of AFM with Outliers}

As pointed out in Remark 1.2, since the consistency of our QFA estimator does not require the moments of the idiosyncratic errors to
exist, at $\tau =0.5$ it can be viewed as a robust QR alternative to the PCA estimators commonly used in practice. For the same token, our estimator of the number of factors should
also be more robust to outliers and heavy tails than the
IC-based method of \textcolor{blue}{BN (2002)}. In this subsection we confirm the above claims by means of Monte Carlo simulations.

We consider the following DGP:
\begin{equation*}
X_{it}=\sum_{j=1}^{3}\lambda _{ji}f_{jt}+u_{it},
\end{equation*}%
where $f_{1t}=0.8f_{1,t-1}+\epsilon _{1t}$, $f_{2t}=0.5f_{2,t-1}+\epsilon
_{2t}$, $f_{3t}=0.2f_{3,t-1}+\epsilon _{3t}$, $\lambda _{ji},\epsilon _{jt}$
are all independent draws from $\mathcal{N}(0,1)$, and $u_{it}\sim i.i.d \text{ }B_{it}\cdot \mathcal{N}(0,1) + (1-B_{it})\cdot\text{Cauchy}(0,1)$, where $B_{it}$ are i.i.d Bernoulli random variables with means equal to $0.98$ and $\text{Cauchy}(0,1)$ denotes the standard Cauchy distribution. In this way, approximately $2\%$ of the idiosyncratic errors are generated as outliers.  


We consider four
estimators of the number of factors $r$: two estimators based on $PC_{p1}$, $%
IC_{p1}$ of \textcolor{blue}{BN (2002)}, the Eigenvalue Ratio (ER) estimator proposed by %
\citet{ahn2013eigenvalue} and our rank-minimization estimator discussed in subsection 3.2,
having chosen
\begin{equation*}
P_{NT}=\hat{\sigma}^k_{N,1}\cdot \left( \frac{1}{L_{NT}^{2}}\right) ^{1/3}.
\end{equation*}%
We set $k=8$ for all four estimators, and consider $N,T\in
\{50,100,200,500\}$.

Table 1 reports the following fractions:
\begin{equation*}
\lbrack \text{proportion of $\hat{r}<3$ },\text{ proportion of $\hat{r}=3$ },%
\text{ proportion of $\hat{r}>3$ }]
\end{equation*}%
for each estimator having run 1000 replications.

It becomes evident from the results in Table 1 that $PC_{p1}$ and $IC_{p1}$ almost always overestimate the number factors, while the ER estimator tends to underestimate them, though to a lesser extent than what $PC_{p1}$ and $IC_{p1}$ overestimate them. By contrast, our rank-minimization estimator chooses accurately the right number of factors as long as $\min \{N,T\}\geq 100$.

Next, to compare the PCA and QFA estimators of the common factors in the previous DGP, we assume that $r=3$
is known. We first get the PCA estimator (denoted as $\hat{F}_{PCA}$), and then obtain the QFA estimator at $\tau=0.5$ (denoted $\hat{F}_{QFA}^{0.5}$) using the IQR algorithm. Next, we regress each of the true factors on $\hat{F}_{PCA}$ and $\hat{F}_{QFA}^{0.5}$ separately, and report the average $R^{2}$ from 1000 replications in Table 2 as an indicator of how well the space of the true factors is spanned by the
estimated factors.\footnote{All the $R^2$ we use in this section and the next section are adjusted $R^2$.} As shown in the first three columns of Table 2, while the PCA estimators are not very successful in capturing the true common factors, the QFA estimators approximate them very well, even when $N,T$ are not too large.

As discussed earlier, the overall findings reported in Tables 1 and 2 are in line with our theoretical results. In effect, while the standard PCA estimators of \textcolor{blue}{BN (2002)} fail to capture the true factors because they require the eighth moments of all the idiosyncratic errors to be bounded (unlike the DGP above), our QFA estimators succeed to do so since they only need the density function to exist and be continuously differentiable, like in the previous DGP. Thus, this simulation exercise provides strong evidence about the substantial gains that can be achieved by using QFA rather than PCA in those cases where the idiosyncratic error terms in AFM exhibit heavy tails and outliers. 

%

\subsection{Estimation of QFM: Heavy-tailed and Dependent Idiosyncratic Errors}

In this subsection we consider the following DGP:
\begin{equation*}
X_{it}=\lambda _{1i}f_{1t}+\lambda _{2i}f_{2t}+(\lambda _{3i}f_{3t})\cdot
e_{it},
\end{equation*}%
where $f_{1t}=0.8f_{1,t-1}+\epsilon _{1t}$, $f_{2t}=0.5f_{2,t-1}+\epsilon
_{2t}$, $f_{3t}=|g_{t}|$, $\lambda _{1i},\lambda _{2i},\epsilon
_{1t},\epsilon _{2t},g_{t}$ are all independent draws from $\mathcal{N}(0,1)$%
, and $\lambda _{3i}$ are independent draws from $U[1,2]$. Following %
\textcolor{blue}{BN (2002)}, the following specification for $e_{it}$ is
used:
\begin{equation*}
e_{it}=\beta \ e_{i,t-1}+v_{it}+\rho \cdot\ \sum_{j=i-J,j\neq
i}^{i+J}v_{jt},
\end{equation*}%
where $v_{it}$ are independent draws from $\mathcal{N}(0,1)$ except in the second case below. The autoregressive coefficient  $\beta $ captures the serial correlation of $e_{it}$, while the parameters $\rho $ and $J$ capture the cross-sectional correlations of $%
e_{it}$. We consider four cases:

\begin{enumerate}
\item[Case 1:] Independent errors: $\beta=0$ and $\rho=0$.

\item[Case 2:] Independent errors with heavy tails: $\beta=\rho=0$, and $%
v_{it}\sim i.i.d\text{ Student}(3)$.

\item[Case 3:] Serially correlated errors: $\beta=0.2$ and $\rho=0$.

\item[Case 4:] Serially and cross-sectionally correlated errors: $\beta=0.2$
and $\rho=0.2$, and $J=3$.
\end{enumerate}

For each of the previous cases and each $\tau \in \{0.25,0.5,0.75\}$, we first
estimate $\hat{r}$ using our rank-minimization estimator, having set $k$ and
$P_{NT}$ as described in the previous subsection. Second, we estimate $\hat{r%
}$ factors by means of the QFA estimation approach, which we denote $\hat{F}%
_{QFA}^{\tau}$. Finally, we regress each of the true factors on $\hat{F}%
_{QFA}^{\tau}$ and calculate the $R^{2}$s. This procedure is repeated
1000 times where, for each $\tau$, we report the averages of $\hat{r}$ and the
$R^{2}$s in these 1000 replications.

The results for Case 1 and Case 2 (where the heavy tails are captured this time by a Student(3) rather than by a Cauchy distribution) are reported in Table 3 and Table 4, respectively, for $N,T\in \{50,100,200\}$. Notice that for $\tau =0.25, 0.75$, we have $r(\tau )=3$ while, for $\tau =0.5$, we get $r(\tau )=2$, since the
factor $f_{3t}$ does not affect the median of $X_{it}$. It can be observed that both our selection criterion and the  QFA estimators perform very well in choosing the number of QFA factors and in estimating them. It should be noticed that at $\tau =0.25,0.75$
the estimation of the scale factor $f_{3t}$ is not as good as the mean factors $f_{1t},f_{2t}$ for small $N$ and $T$. However, such differences vanish as $N$ and $T$ increase.

The results for Case 3 and Case 4 are in turn reported in Table 5 and Table 6, respectively. It can be inspected that the QFA estimators still perform well, even though the independence assumption is violated in these DGPs. Thus, despite adopting independence in Assumption 1 (iii) for tractability in the proofs (see Remark 1.4), it seems that QFA estimation still works properly when the error terms are allowed to exhibit mild serial and cross-sectional correlations.

\subsection{Normal Approximations of the Estimators Based on SQR}
To evaluate the normal approximations of Theorem 4 for the estimators based on SQR, we consider the following DGP:
\[X_{it} = \lambda_i f_t + f_t \epsilon_{it},\]
where $f_t \sim i.i.d \text{ }\mathcal{U}(1,2)$ and they are normalized such that $F'F/T=1$, $\lambda_{i} \sim i.i.d \text{ }\mathcal{N}(0,1)$ and $\epsilon_{it} \sim i.i.d \text{ }\mathcal{N}(0,1)$. Note that since our Theorem 4 is conditional on the factors and the loadings, $f_t$ and $\lambda_i$ are fixed in the simulations. To smooth the indicator function, we use the following eighth-order kernel function (see \citealp{muller1984smooth}):
\[ k(z) = \mathbf{1}\{|z|\leq 1\} \cdot \frac{3465}{8192} \left( 7 - 105z^2 + 462 z^4 -858 z^6 + 715 z^8 -221 z^{10}\right), \]
while the Epanechnikov kernel $l(z) = 0.75(1-z^2)\cdot \mathbf{1}\{|z|\leq 1\} $ is applied to estimate the variance.

Figure 1 and Figure 2 plot the histograms of the standardized estimators of the factors: $\hat{V}_{f_t}^{-1}\sqrt{N}(\tilde{f}_t -f_{0t})$ at $\tau=0.25$, $t=T/2$ from 1000 replications\footnote{We choose the signs of $\tilde{f}_t$ such that $\Sz=1$.}, where $\hat{V}_{f_t}$ is estimated using the formula in Remark 4.4. To check how the bandwidths affect the finite distributions of the estimators, we display results for different choices of $h$ and $b$.

From the reported histograms of the standardized estimators and the superimposed density function of the standard normal distribution, it becomes clear that the asymptotic distributions given in Theorem 4 provide reasonably good approximations for the finite sample distributions of the estimators based on SQR, even for $N=T=50$, and that such approximations are not very sensitive to the choice of bandwidths.   

\section{Empirical Applications}
In this section we argue that QFA could provide a useful tool for causal analysis, predictive exercises and the economic interpretation of factors. In particular, we focus on applying our proposed methodology to three different datasets related to climate, macro aggregates and stock returns. 
\subsection{Climate Change and $\text{CO}_2$ Emissions}
In our first empirical application we investigate how $\text{CO}_2$ emissions affect temperatures, which is a long-standing issue in climate change science and economics (see e.g. \citealt{hansen1981climate}, and \citealt{hsiang2018economist}). The dataset (coined Climate for short) we use consists of the annual changes
of temperature from 441 stations from 1917 to 2018 ($N=441,T=102$), drawn
from the Climate Research Unit at the University of East Anglia, where
information about global temperatures across different stations in the
Northern and Southern Hemisphere is collected. The annual global $\text{CO}_2$ emissions data is downloaded from The Global Change Data Lab.

Table 7 (column labeled Climate) reports the estimated number of factors using $PC_{p1}$ of \textcolor{blue}{BN (2002)}, the ER estimator and the rank-minimization estimator for a grid of quantiles ranging from 0.01 to 0.99.\footnote{In all the applications, before estimating the factors and the number of factors, each variable is standardized to have zero mean and variance equal to one.} The maximum number of factors $k$ is set to $8$ for all estimators. $PC_{p1}$ selects the maximum number of factors (8), while the ER estimator selects only one. Thus, the tendencies to overestimate (resp. underestimate)  the number of PCA factors by the former (resp. latter) criteria mirror our simulation results in Section 5.1. By contrast, it can be observed that the numbers of factors estimated by the rank-minimization vary across quantiles. In particular, the number of QFA factors decreases as we move away from the median.

To compare the QFA factors (denoted as $\hat{F}_{QFA}^{\tau}$) and the PCA factors (denoted as $\hat{F}_{PCA}$), we regress each element of $\hat{F}_{QFA}^{\tau}$ on the 8 PCA factors selected by $PC_{p1}$ and compute the $R^2$ in these regressions.\footnote{We choose the number of PCA factors estimated by $PC_{p1}$ in these regressions to play conservative.} The results are shown in the upper panel of Table 8, where it becomes clear that the median factors ($\hat{F}_{QFA}^{0.5}$) are highly correlated with the PCA factors, with all the $R^2$s above $0.95$. By contrast,  the QFA factors at the upper and lower quantiles ($\tau=0.01,0.05,0.95,0.99$) exhibit much lower correlations with the PCA factors, with $R^2$s around $0.60$. Thus, there seems to be room for using QFA in this application. 

Next, to analyze the impact of $\text{CO}_2$ emissions on climate change, bivariate Granger non-causality tests are implemented. We regress the QFA factors at each relevant quantile on their own lags and the lagged growth rates of $\text{CO}_2$ emissions, labeled $\Delta \log ( \text{CO}_2)$, where the lag length is chosen according to BIC. Table 9 reports the p-values of these tests. The results of this novel approach to analyze quantile causality indicate that the growth rate of $\text{CO}_2$ emissions strongly Granger causes the QFA factors at the lower quantiles ($\tau=0.01,0.05$), with p-values below 0.01, as well as some of the median factors, albait to a lessser extent (p-values below 0.04). Moreover, the null of Granger non-causality is not rejected for QFA factors at the upper quantiles. Given that $\text{CO}_2$ emissions lead to global warming, the results for the lower quantiles of temperatures are in line with the evidence reported by \citet{rivas2020trends}. Using a similar climate dataset but different quantile techniques to ours, these authors find that global warming over the last century seems to be mainly due to a different behaviour in the lower tail than in the central and upper tails of the distribution of global temperatures.

\subsection{Macroeconomic Forecasting in a Data-Rich Environment}

In the second application, we extend the diffusion-index forecasting exercise popularized by \citet{stock2002forecasting} to explore the predictive power of the QFA factors. The main goal is to extract a few common factors (by both PCA and QFA) from a large panel of macroeconomic variables, and then use these factors to forecast e.g. real GDP growth and the inflation rate. 

The FRED-QD dataset (coined Macro here) is used to estimate PCA and QFA factors. This is a quarterly panel consists of 211 US macroeconomic variables from 1960Q1 to 2019Q2 ($N=211, T=238$). It emulates the popular dataset used by \citet{stock2002forecasting}, but also contains several additional time series. The variables in this dataset are updated in a timely manner and can be downloaded for free.\footnote{Link to the dataset: http://research.stlouisfed.org/econ/mccracken/. We refer to \citet{mccracken2016fred} for the details of a very similar dataset that contains monthly macroeconomic variables.} Before estimation, each series is transformed to be stationary using Matlab codes that are also available on the FRED-QD data website.  

As with the climate data, while $PC_{p1}$ selects 8 PCA factors, ER selects only one. The estimated numbers of QFA factors are reported in Table 7 (column labeled Macro). As can be seen, the number of QFA factors varies significantly across different quantiles, pointing to the existence of a nonstandard factor structure for this dataset. Moreover, the middle panel of Table 8 reports the $R^2$s of regressing each of the QFA factors on the 8 PCA factors. It becomes clear that the QFA factors at $\tau$ close to $0.5$ are all well explained by the PCA factors. However, the first QFA factor at $\tau=0.9$ (denoted $\hat{F}_{QFA}^{0.9}$) and those at $\tau=0.95,0.99$ (denoted as $\hat{F}_{QFA}^{0.95}$ and $\hat{F}_{QFA}^{0.99}$) contain some extra information that could be potentially helpful for forecasting macroeconomic variables. Since $\hat{F}_{QFA}^{0.95}$ exhibits a very high correlation with $\hat{F}_{QFA}^{0.9}$ and $\hat{F}_{QFA}^{0.99}$, we exclusively focus on the predictive power of $\hat{F}_{QFA}^{0.9}$ and $\hat{F}_{QFA}^{0.99}$ in the subsequent analysis.

Let $y_{t+1}$ denote the realized value of real GDP growth/inflation at period $t+1$. The forecasting equation we consider is as follows:
\[ y_{t+1} = \alpha + \sum_{j=0}^{p_{max}} \beta_j y_{t-j} + \gamma' F_t + \epsilon_{t+1},\] 
where $F_t$ is vector containing several unobserved common factors extracted from the large macroeconomic dataset. The predicted value of $y_{t+1}$, based on a vector of estimated factors $\hat{F}_t$, is simply constructed as $\hat{y}_{t+1} = \hat{\alpha}+  \sum_{j=0}^{\hat{p}} \hat{\beta}_{j} y_{t-j} + \hat{\gamma}' \hat{F}_t $, where $ \hat{\alpha},\hat{\beta}_{j},$ $\hat{\gamma}$ are OLS estimates of the coefficients and $\hat{p}$ is the optimal lag length according to BIC. We compare five different specifications for $F_t$: (i) $F_t=0$, which is the benchmark AR model, (ii) AR plus $\hat{F}_t$ only including $\hat{F}_{PCA}$, (iii) AR plus $\hat{F}_t$ including $\hat{F}_{PCA}$ and $\hat{F}_{QFA}^{0.9}$, (iv) AR plus $\hat{F}_t$ including $\hat{F}_{PCA}$ and $\hat{F}_{QFA}^{0.99}$, and (v) AR plus $\hat{F}_t$ including $\hat{F}_{PCA}$, $\hat{F}_{QFA}^{0.9}$ and $\hat{F}_{QFA}^{0.99}$. Following \citet{chudik2018one}, the initial estimation period is 1960Q1 to 1989Q4 (120 periods), and the forecast evaluation period is split into pre-crisis  (1990Q1 to 2007Q2) and crisis and recovery (2007Q3 to 2019Q2) sub-periods. A rolling window of 120 periods is used both to estimate the coefficients and generate the rolling forecasts. In particular, following \citet{chudik2018one}, the number of mean factors is estimated using $PC_{p1}$ at each rolling window, where the maximum number of factors is set equal to 5.

The mean squared error (MSE) of these procedures, and their relative MSE (R-MSE) to the benchmark AR model are reported in Table 10 for the whole evaluation period and each relevant sub-sample. As can be observed, in regards to real GDP growth, adding the upper tail QFA factors ranks better in terms of R-MSE than the AR and AR+$\hat{F}_{PCA}$ models for the three considered periods. The gains are not sizeable but yet they are relevant. As for the inflation rate, the results are weaker, though there are some gains for the crisis and recovery period.

A well-known shortcoming of point forecasts is that their uncertainty is generally unknown, hence it is difficult to quantify their precision at any given period of time. To address this problem, it has became customary among  central banks to report density forecasts for important macroeconomic variables. In this respect,  \citet{adrian2019vulnerable} argue that a simple way of producing such densities is via QR. Following these authors, we next evaluate the predictive power of the QFA factors for forecasting the densities of real GDP growth and inflation. In particular, we first predict the conditional quantiles of the target variable $y_{t+h}$ by $\hat{q}_{\tau,t+h} = \hat{\alpha}_{\tau}+ \sum_{j=0}^{p} \hat{\beta}_{\tau,j} y_{t-j} +\hat{\gamma}_{\tau}' \hat{F}_{\tau,t}$ for $\tau\in\{0.05,0.25,0.75,0.95\}$, where $ \hat{\alpha}_{\tau},\hat{\beta}_{\tau,j},\hat{\gamma}_{\tau}$ are estimated coefficients by running QR of $y_{t+h}$ on $[1,y_t,\ldots, y_{t-p}, \hat{F}_{\tau,t}]$, and $\hat{F}_{\tau,t}$ is a vector of estimated quantile factors using the IQR algorithm.\footnote{Only 1 QFA factor is estimated at $\tau=0.05, 0.95$, whereas 5 QFA factors are estimated at $\tau=0.25, 0.75$.} Next, given the predicted quantiles: $[\hat{q}_{0.05,t+h}, \hat{q}_{0.25,t+h},\hat{q}_{0.75,t+h},\hat{q}_{0.95,t+h} ]$, the predicted density of $y_{t+h}$ is constructed as the density of a skewed $t$-distribution by matching the predicted quantiles.\footnote{We refer to \citet{adrian2019vulnerable} for the details and to \citet{azzalini2003distributions} for the definition and properties of the skewed $t$-distribution.} Finally, the accuracy of the density forecast is measured by the predictive score, which is the predicted density evaluated at the realized value of $y_{t+h}$. Higher predictive scores indicate more accurate predictions. The out-of-sample density forecasts are constructed using rolling windows with the most recent 120 observations, and the evaluation period is 1990Q1 to 2019Q2. Moreover, we set $p=3$, and the benchmark model is the one where $\hat{F}_{\tau,t}=0$, i.e. the quantiles of $y_{t+h}$ are predicted only using its own lags. Figure 3 displays the predictive scores of the one-quarter-ahead ($h=1$) and one-year-ahead ($h=4$) density forecasts for both variables. It can be seen that in both instances the predictive scores of the``AR + Quantile Factors'' procedure is frequently above that of the ``AR benchmark'' model,  sometimes by a large margin, indicating that the QFA factors could indeed be very informative for density forecasting of highly relevant macroeconomic variables.

\subsection{Interpretation of Financial Factors}
Our last application concerns the interpretation of the quantile factors extracted from financial asset returns. The dataset (Finance in short) contains the monthly returns of 429 stocks from 1980M01 to 2014M12 ($N=429, T=420$), obtained from The Center of Research for Security Prices (CRSP).\footnote{The panel is balanced by only keeping stocks that have no missing observations during this time period. }

Except at $\tau=0.5$, the estimated number of QFA factors reported in Table 7 (column labeled Finance) are all equal to 1, which agrees with the choice of PCA factors by the ER estimator but again is less than the 4 factors selected by $PC_{p1}$.\footnote{$PC_{p3}$ and $IC_{p3}$ of \textcolor{blue}{BN (2002)} chose 8 factors while all the other 6 information criteria choose 4 factors.} 

The lower panel of Table 8 reports the $R^2$s of regressing each of the QFA factors on the 8 PCA factors. As can be inspected, most of these factors are well explained by the PCA factors, with the exception of those at $\tau=0.01, 0.99$, where the $R^2$s are below $60\%$. Interestingly, as discussed in the Introduction this evidence is seemingly consistent with the findings of the financial literature on the existence of tail factors in the distribution of asset returns, as reported e.g. by  \citet{RePEc:aah:create:2018-02}. Thus, it is interesting to check  whether the extra quantile factors at the lower tail and upper tail of the returns distribution could yield some confirmation of that hypothesis.

First, as shown in the upper panel of Figure 4, the QFA factors at $\tau=0.01$ and 0.99 (both with variance $=0.12$) are much less volatile than those at $\tau=0.5$ (both with variance $=1$), meaning that the tails of the distributions of returns are more stable than the median. Second, we find that the interquantile range (defined as the difference between the quantile factors at $\tau=0.99$ and $0.01$) provides
a good measure of uncertainty for financial markets.\footnote{The results with the interpercentile range and interquartile range turn out to be similar.} Finally, as shown in the lower panel of Figure 4, the interquantile range is highly correlated with the volatility factor (with a correlation of 0.87) constructed by applying PCA-SQ to the squared residuals of an AFM.\footnote{ Following \citet{renault2016apt}, we first project out the 8 PCA factors from the returns, and the volatility factor is obtained as the cross-sectional average of the squared residuals.} On the contrary, the correlations between the two median factors and the volatility factor only reach 0.08 and -0.05, respectively. Thus, this evidence seems supportive of the the presence of extra common factors affecting the tails and the volatility of the asset returns, with our results providing a link between them.

\section{Conclusions}
Approximate Factor Models (AFM) have become a leading methodology for the joint modelling of large number of economic time series with the big improvements in data collection and information technologies. This first generation of AFM was designed to reduce the dimensionality of big datasets through finding those common components (mean factors) which, by shifting the means of the observed variables with different intensities, are able to capture a large fraction of their co-movements. However, one could envisage the existence of other common factors that do not (or not only) shift the means but also affect other distributional characteristics (volatility, higher moments, extreme values, etc.). This calls for a second generation of factor models.

Inspired by the generalization of linear regressions to quantile regressions (QR), this paper proposes Quantile Factor Models (QFM) as a new class of factor models. In QFM, both factors and loadings are allowed to be quantile-dependent objects. These extra factors could be useful for identification purposes, for instance mean factors vs. volatility/skewness/kurtosis factors, as well as for forecasting purposes in factor-augmented regressions and FAVAR setups.

Using tools in the interface of QR, Principal Component Analysis (PCA) and the theory of empirical processes, we propose an estimation procedure of the quantile-dependent objects in QFM, labelled Quantile Factor Analysis (QFA), which yields consistent and asymptotically normal estimators of factors and loadings at each quantile. An important advantage of QFA is that it is able to extract simultaneously all mean and extra (non-mean) factors determining the factor structure of QFM, in contrast to PCA which can only extract mean factors. In addition, we propose novel selection criteria to estimate consistently the number of factors at each quantile. Finally, another relevant result is that QFA estimators remain valid (under some restrictive assumption on the idiosyncratic error terms -- see Assumption1 (iii) -- which are adopted to simplify the proofs) when the idiosyncratic error terms in AFM exhibit heavy tails and outliers, a case where PCA is rendered invalid.

The previous theoretical findings receive support in finite samples from a range of Monte Carlo simulations. Furthermore, it is shown in these simulations that QFA estimation performs well when we depart from some of simplifying assumptions used in the theory section for tractability, like, e.g., independence of the idiosyncratic errors. Lastly, our empirical applications to three large panel datasets of financial, macro and climate variables provide evidence that some of these extra factors may be highly relevant in practice for causality analysis, forecasting, and economic interpretation purposes.

Any time a novel methodology is proposed, new research issues emerge for future investigation. Among the ones which have been left out of this paper (some are part of our current research agenda), four topics stand out as important:
\begin{itemize}
    \item Factor augmented regressions and FAVAR:
    In relation to this topic, it would also be interesting to check in great detail the contributions of the extra factors in forecasting and monitoring multivariate systems. This is an issue of high interest for applied researchers, especially with the surge of Big Data technologies. For example, one could analyze the role of the extra factors in the estimation and shock identification in FAVAR. Recent developments in quantile VAR estimation, as in \citet{white2015var}, provide useful tools in addressing these issues.


    \item Relaxing the independence assumptions: In view of  the simulation results in Tables 5 and 6, we conjecture that the main theoretical results of our paper continue to hold when the error terms in QFM are allowed to have weak cross-sectional and serial dependence. Providing a formal justification for this conjecture remains high in our research agenda. As discussed in Remark 1.4, the goal here is to provide more general conditions on $u_{it}$ under which the sub-Gaussian type inequalities still hold.

    \item Dynamic QFM: Although our methodology admits  factors to have dependence, provided Assumption 2(i) holds, there is still the pending issue of how to extend our results for static QFM to dynamic QFM, where the set of quantile-dependent variables include lagged factors (see \citealt{forni2000generalized} and \citealt{stock2011dynamic}).  Since our main aim in this paper has been to introduce the new class of QFM and their basic properties, for the sake of brevity, we have focused on static QFM, leaving this topic for further research.

    \item Economic interpretation of QFA factors in empirical applications: Given the evidence that extra factors could be relevant in practice, another interesting issue is how to interpret them in different economic and financial setups. As illustrated in subsection 6.3, once the econometric techniques to detect and estimate extra factors in QFM have been established, attempts to provide new economic insights for these objects would help enrich the economic theory underlying this type of factor structures.

\end{itemize}

\newpage \appendix

\section{Tables and Figures}

\begin{table}[htp]
\begin{center}
\begin{threeparttable}
\caption{AFM with Outliers in the Idiosyncratic Errors: Estimating the Number of Factors}
\begin{tabular}{cccccccccccccc}
\hline
$N$ & $T$   &       \multicolumn{3}{c}{ $PC_{p1}$ of BN}  &     \multicolumn{3}{c}{ $IC_{p1}$ of BN }     &  \multicolumn{3}{c}{Eigenvalue Ratio} &  \multicolumn{3}{c}{Rank Estimator} \\
\cmidrule(lr){1-2}   \cmidrule(lr){3-5} \cmidrule(lr){6-8}  \cmidrule(lr){9-11} \cmidrule(lr){12-14}
 50&50 &  [0.00&0.04 &0.96] &[0.00  &0.14&0.86] &[0.26&0.30&0.44]& [0.47 &0.53&0.00] \\
 50&100 &[0.00 & 0.02&0.98] &[0.00 &0.05& 0.95]&[0.33&0.19&0.48]& [0.40&0.60& 0.00]\\
 50&200 &[0.00 & 0.00&1.00] &[0.00  &0.01&0.99] &[0.41&0.12&0.47]& [0.33&0.67&0.00] \\
 50&500 &[0.00 & 0.00& 1.00] &[0.00  &0.00& 1.00]&[0.56&0.07&0.37]& [0.29&0.71&0.00] \\
 \hline
 100&50 &[0.00 & 0.02 &0.98] &[0.00  &0.05& 0.95] &[0.34&0.18&0.48]& [0.39&0.61&0.00] \\
 100&100 &[0.00 & 0.00&1.00] &[0.00  &0.01& 0.99]&[0.41&0.13&0.46]& [0.10&0.90&0.00] \\
 100&200 &[0.00 & 0.00&1.00] &[0.00  &0.00& 1.00]&[0.48&0.07&0.45]& [0.06&0.94&0.00] \\
 100&500 &[0.00 & 0.00& 1.00]&[0.00  &0.00& 1.00]&[0.65&0.05&0.30]& [0.02&0.98 &0.00] \\
\hline
 200&50 &[0.00 &0.00 & 1.00]&[0.00  &0.01& 0.99]&[0.45&0.10&0.45]& [0.37&0.63&0.00] \\
 200&100 &[0.00&0.00 &1.00] &[0.00  &0.00& 1.00]&[0.48&0.08&0.44]& [0.10&0.90&0.00] \\
 200&200 &[0.00 &0.00 &1.00] &[0.00  &0.00& 1.00]&[0.63&0.06&0.31]& [0.00&1.00&0.00] \\
 200&500 & [0.00&0.00 &1.00] &[0.00  &0.00& 1.00] &[0.76&0.08&0.16]& [0.00&1.00&0.00] \\
\hline
 500&50 &[0.00 &0.00 & 1.00]&[0.00  &0.00&1.00] &[0.57&0.08&0.35]&[0.36&0.64&0.00] \\
 500&100 &[0.00 &0.00 &1.00] &[0.00  &0.00&1.00] &[0.68&0.06&0.26]& [0.05&0.95&0.00] \\
 500&200 &[0.00 & 0.00& 1.00]&[0.00  &0.00& 1.00]&[0.76&0.08&0.16]& [0.00&1.00&0.00] \\
 500&500 &[0.00 & 0.00&1.00] &[0.00  &0.00& 1.00]&[0.80&0.10&0.10]& [0.00&1.00&0.00] \\
 \hline
\end{tabular}
\begin{tablenotes}
      \small
      \item Note: The DGP considered in this Table is: $X_{it} = \sum_{j=1}^{3} \lambda_{ji} f_{jt} +u_{it}$, where $f_{1t} =0.8 f_{1,t-1}+\epsilon_{1t}$, $f_{2t} =0.5 f_{2,t-1}+\epsilon_{2t}$, $f_{3t} =0.2 f_{3,t-1}+\epsilon_{3t}$, $\lambda_{ji},\epsilon_{jt}\sim i.i.d \text{ }\mathcal{N}(0,1)$, $u_{it}\sim i.i.d \text{ }B_{it}\cdot \mathcal{N}(0,1) + (1-B_{it})\cdot\text{Cauchy}(0,1)$ where $B_{it}\sim i.i.d \text{ Bernoulli}(0.98)$. For each estimation method, the $[ \text{proportion of $\hat{r}<3 $ }, \text{ proportion of $\hat{r}=3 $ },  \text{ proportion of $\hat{r}>3 $ } ]$ is reported from 1000 replications.
    \end{tablenotes}
\end{threeparttable}
\end{center}
\end{table}

\begin{table}[htp]
\begin{center}
\begin{threeparttable}
\caption{AFM with Outliers in the Idiosyncratic Errors: Estimation of the Factors}
\begin{tabular}{cccccccc}
\hline
  & & \multicolumn{3}{c}{Regress $F$ on $\hat{F}_{PCA}$} & \multicolumn{3}{c}{Regress $F$ on $\hat{F}_{QFA}^{0.5}$ }\\
  $N$ &  $T$  &    $f_{1}$     &   $f_{2}$   &      $f_{3}$  &    $f_{1}$   &   $f_{2}$    &   $f_{3} $   \\
\cmidrule(lr){1-2}   \cmidrule(lr){3-5} \cmidrule(lr){6-8}
 50&50& 0.939&0.810 &0.686 & 0.987 &0.975 &0.968\\
 50&100& 0.931 &0.718&0.578 &0.987 & 0.975 & 0.968\\
 50&200&0.890&0.589&0.412 & 0.987&0.975 &0.968\\
 50&500&0.807&0.405&0.252 & 0.988&0.975 &0.968\\
 \hline
 100&50&0.928&0.738&0.595 &0.993 &0.986 &0.984\\
 100&100&0.921&0.630&0.441 & 0.994 &0.988 &0.984\\
 100&200&0.857&0.479&0.285 &0.994 &0.988 &0.985\\
 100&500&0.713&0.294&0.138 &0.994 &0.988 &0.984\\
 \hline
 200&50&0.890&0.657&0.513 & 0.997 &0.994 &0.992\\
 200&100&0.858&0.514&0.333 &0.997 &0.994 & 0.993\\
 200&200&0.779&0.358&0.178 & 0.997 &0.994 &0.992\\
  200&500&0.530&0.131&0.051 & 0.997 &0.994 &0.992\\
 \hline
 500&50&0.819&0.501&0.371 & 0.998 &0.997 &0.996\\
 500&100&0.725&0.327&0.196 &0.999 &0.998 & 0.997\\
 500&200&0.546&0.165&0.062 & 0.999 &0.998 &0.997\\
  500&500&0.273&0.036&0.018 & 0.999 &0.998 &0.997\\
 \hline
\end{tabular}
\begin{tablenotes}
      \small
      \item Note: The DGP considered in this Table is: $X_{it} = \sum_{j=1}^{3} \lambda_{ji} f_{jt} +u_{it}$, where $f_{1t} =0.8 f_{1,t-1}+\epsilon_{1t}$, $f_{2t} =0.5 f_{2,t-1}+\epsilon_{2t}$, $f_{3t} =0.2 f_{3,t-1}+\epsilon_{3t}$, $\lambda_{ji},\epsilon_{jt}\sim i.i.d \text{ }\mathcal{N}(0,1)$, $u_{it}\sim i.i.d \text{ }B_{it}\cdot \mathcal{N}(0,1) + (1-B_{it})\cdot\text{Cauchy}(0,1)$ where $B_{it}\sim i.i.d \text{ Bernoulli}(0.98)$. For each estimation method, we report the average $R^2$ in the regression of (each of) the true factors on the estimated factors by PCA and QFA (assuming the number of factors to be known).
    \end{tablenotes}
\end{threeparttable}
\end{center}
\end{table}

\begin{table}[htp]
\begin{center}
\begin{threeparttable}
\caption{Estimation of QFM: Independent Idiosyncratic Errors}

\begin{tabular}{cc|cccc|cccc|cccc}
\hline
          &         &  \multicolumn{4}{|c|}{ $\tau=0.25$ }    &  \multicolumn{4}{c|}{ $\tau=0.5$ }    &  \multicolumn{4}{c}{ $\tau=0.75$ }  \\
  $N$ &  $T$ &   $\hat{r}$     &   $f_{1t}$   &      $f_{2t}$  &    $f_{3t}$   &  $\hat{r}$     &   $f_{1t}$   &      $f_{2t}$  &    $f_{3t}$ &  $\hat{r}$     &   $f_{1t}$   &      $f_{2t}$  &    $f_{3t}$     \\
\hline
 50&50&2.21 &0.866&0.721 &0.339 & 1.91  & 0.956 &0.808 &0.013  & 2.23& 0.926 &0.738 &0.334 \\
 50&100& 2.42&0.943 &0.758&0.483 & 1.88&0.968 & 0.839 & 0.003 & 2.38& 0.946 &0.708 &0.463\\
  50&200& 2.43&0.933&0.703&0.485 & 1.88&0.971&0.842 &0.001 & 2.40&0.951 &0.698 &0.445\\
\hline
 100&50& 2.14&0.944&0.681 &0.337 & 1.80 & 0.980 &0.786&0.014  & 2.13& 0.948 &0.694 &0.357 \\
 100&100& 2.71&0.977 &0.898&0.688 & 1.98&0.985 & 0.954 & 0.001 & 2.72& 0.968 &0.890 &0.707\\
  100&200& 2.82&0.983&0.904&0.757 & 1.99& 0.987&0.966 &0.003 & 2.86&0.982 &0.908 &0.793\\
  \hline
 200&50& 2.35&0.970&0.826 &0.490 & 1.87& 0.989 &0.867 &0.008  & 2.29& 0.973 &0.745 &0.489 \\
 200&100& 2.80&0.990 &0.934&0.782 &2.00&0.993 & 0.987 & 0.001 & 2.81& 0.990 &0.977 &0.772\\
  200&200&2.99 &0.992&0.986&0.940 & 2.00&0.994&0.988 &0.000 & 2.99&0.992 &0.986 &0.935\\
  \hline
\end{tabular}
\begin{tablenotes}
      \small
      \item Note: The DGP considered in this Table is: $X_{it} = \lambda_{1i} f_{1t} +  \lambda_{2i} f_{2t} + ( \lambda_{3i} f_{3t}) \cdot e_{it}$, $f_{1t} =0.8 f_{1,t-1}+\epsilon_{1t}$, $f_{2t} =0.5 f_{2,t-1}+\epsilon_{2t}$, $f_{3t} =|g_t|$, $\lambda_{1i},\lambda_{2i},\epsilon_{1t},\epsilon_{2t},g_t \sim i.i.d \text{ }\mathcal{N}(0,1)$, and $\lambda_{3i} \sim i.i.d \text{ }U[1,2]$. $e_{it} = \beta e_{i,t-1} + v_{it} + \rho \cdot \sum_{j=i-J,j\neq i}^{i+J}v_{jt}$, $v_{it}\sim i.i.d \text{ }\mathcal{N}(0,1)$, $\beta=\rho=0$. For each $\tau$, the first column reports the averages of the rank estimator $\hat{r}$ from 1000 replications, while the second to the fourth columns report the average $R^2$ in the regression of (each of) the true factors on the QFA factors $\hat{F}^{\tau}_{QFA}$, obtained from the IQR algorithm.
    \end{tablenotes}
\end{threeparttable}
\end{center}
\end{table}

\begin{table}[htp]
\begin{center}
\begin{threeparttable}
\caption{Estimation of QFM: Independent Idiosyncratic Errors with Heavy Tails}

\begin{tabular}{cc|cccc|cccc|cccc}
\hline
          &         &  \multicolumn{4}{|c|}{ $\tau=0.25$ }    &  \multicolumn{4}{c|}{ $\tau=0.5$ }    &  \multicolumn{4}{c}{ $\tau=0.75$ }  \\
  $N$ &  $T$ &   $\hat{r}$     &   $f_{1t}$   &      $f_{2t}$  &    $f_{3t}$   &  $\hat{r}$     &   $f_{1t}$   &      $f_{2t}$  &    $f_{3t}$ &  $\hat{r}$     &   $f_{1t}$   &      $f_{2t}$  &    $f_{3t}$     \\
\hline
 50&50&2.81 &0.911&0.727 &0.585 & 2.38  & 0.954 &0.827 &0.031  & 2.95 & 0.925 &0.711 &0.617 \\
 50&100& 2.79&0.934 &0.782&0.621 & 2.03&0.963 & 0.885 & 0.005 & 2.79& 0.933&0.783 &0.658\\
  50&200& 2.82 & 0.942& 0.811&0.680 &1.91  &0.966  &0.855 &0.000  & 2.76 &0.943  &0.790 &0.648 \\
\hline
 100&50&3.20 & 0.962&0.851 &0.737 & 2.67  & 0.977  &0.907 & 0.076 &3.07 & 0.942  &0.828 & 0.682 \\
 100&100& 3.06&0.972 &0.897&0.840 & 2.21 &0.983 & 0.939 & 0.018 & 3.06& 0.974 &0.931 &0.801\\
 100&200&3.00 & 0.974& 0.944&0.867 & 1.99  & 0.983 & 0.958& 0.000 & 2.98 &0.974  &0.943 &0.860 \\
  \hline
 200&50& 3.24& 0.971& 0.839& 0.753&2.82  & 0.984  &0.903 & 0.106 &3.31 &0.970  &0.858 &0.773 \\
 200&100& 3.10&0.985& 0.937& 0.897& 2.31 & 0.991  &0.975 & 0.018 & 3.09 & 0.987  & 0.949 &0.883 \\
  200&200&3.02 & 0.989& 0.977& 0.932 & 2.07  & 0.992  & 0.985 & 0.005  &  3.02& 0.988 & 0.978 &0.933 \\
  \hline
\end{tabular}
\begin{tablenotes}
      \small
      \item Note: The DGP considered in this Table is: $X_{it} = \lambda_{1i} f_{1t} +  \lambda_{2i} f_{2t} + ( \lambda_{3i} f_{3t}) \cdot e_{it}$, $f_{1t} =0.8 f_{1,t-1}+\epsilon_{1t}$, $f_{2t} =0.5 f_{2,t-1}+\epsilon_{2t}$, $f_{3t} =|g_t|$, $\lambda_{1i},\lambda_{2i},\epsilon_{1t},\epsilon_{2t},g_t \sim i.i.d \text{ }\mathcal{N}(0,1)$, and $\lambda_{3i} \sim i.i.d \text{ }U[1,2]$. $e_{it} = \beta e_{i,t-1} + v_{it} + \rho \cdot \sum_{j=i-J,j\neq i}^{i+J}v_{jt}$, $v_{it}\sim i.i.d\text{ Student}(3)$, $\beta=\rho=0$. For each $\tau$, the first column reports the averages of the rank estimator $\hat{r}$ from 1000 replications, while the second to the fourth columns report the averages of $R^2$ in the regression of (each of) the true factors on the QFA factors $\hat{F}^{\tau}_{QFA}$, obtained from the IQR algorithm.
    \end{tablenotes}
\end{threeparttable}
\end{center}
\end{table}

\begin{table}[htp]
\begin{center}
\begin{threeparttable}
\caption{Estimation of QFM: Serially Correlated Idiosyncratic Errors}
\begin{tabular}{cc|cccc|cccc|cccc}
\hline
          &         &  \multicolumn{4}{|c|}{ $\tau=0.25$ }    &  \multicolumn{4}{c|}{ $\tau=0.5$ }    &  \multicolumn{4}{c}{ $\tau=0.75$ }  \\
  $N$ &  $T$ &   $\hat{r}$     &   $f_{1t}$   &      $f_{2t}$  &    $f_{3t}$   &  $\hat{r}$     &   $f_{1t}$   &      $f_{2t}$  &    $f_{3t}$ &  $\hat{r}$     &   $f_{1t}$   &      $f_{2t}$  &    $f_{3t}$     \\
\hline
 50&50&2.31 &0.900&0.698 &0.400 & 1.97& 0.961 &0.805 &0.023  & 2.32& 0.924 &0.705 &0.416 \\
 50&100&2.40 &0.927 &0.722&0.475 &1.91&0.968 & 0.863 & 0.005 &2.38& 0.940 &0.709 &0.453\\
  50&200& 2.66&0.956&0.841&0.586 & 1.95&0.970&0.904&0.000 & 2.70&0.948 &0.824 &0.628\\
\hline
 100&50& 2.33&0.945&0.736 &0.479 & 1.91& 0.980 &0.857 &0.005  & 2.32& 0.942 &0.737 &0.478 \\
 100&100& 2.72&0.978 &0.863&0.704 &1.98&0.985 & 0.957 & 0.000 & 2.72& 0.978 &0.895 &0.690\\
  100&200& 2.87&0.983&0.924&0.801 & 1.98&0.987 &0.955 &0.000 & 2.88&0.965 &0.948 &0.805\\
  \hline
200&50&2.35 &0.974&0.724 &0.540 & 1.92 & 0.989 &0.859 &0.021  & 2.40& 0.963 &0.758 &0.531 \\
 200&100& 2.75&0.987 &0.929&0.734 &1.98&0.993 & 0.960 & 0.000 &2.76& 0.990 &0.912 &0.760\\
  200&200&2.98 &0.993&0.984&0.927 & 2.00&0.994&0.987 &0.000 & 2.99&0.992 &0.975 &0.942\\
  \hline
\end{tabular}
\begin{tablenotes}
      \small
      \item Note: The DGP considered in this Table is: $X_{it} = \lambda_{1i} f_{1t} +  \lambda_{2i} f_{2t} + ( \lambda_{3i} f_{3t}) \cdot e_{it}$, $f_{1t} =0.8 f_{1,t-1}+\epsilon_{1t}$, $f_{2t} =0.5 f_{2,t-1}+\epsilon_{2t}$, $f_{3t} =|g_t|$, $\lambda_{1i},\lambda_{2i},\epsilon_{1t},\epsilon_{2t},g_t \sim i.i.d \text{ }\mathcal{N}(0,1)$, and $\lambda_{3i} \sim i.i.d \text{ }U[1,2]$. $e_{it} = \beta * e_{i,t-1} + v_{it} + \rho \cdot \sum_{j=i-J,j\neq i}^{i+J}v_{jt}$, $v_{it}\sim i.i.d \text{ }\mathcal{N}(0,1)$, $\beta=0.2$, $\rho=0$. For each $\tau$, the first column reports the average rank estimator $\hat{r}$ from 1000 replications, while the second to the fourth columns report the average $R^2$ in the regression of (each of) the true factors on the QFA factors $\hat{F}^{\tau}_{QFA}$, obtained from the IQR algorithm.
          \end{tablenotes}
\end{threeparttable}
\end{center}
\end{table}

\begin{table}[htp]
\begin{center}
\begin{threeparttable}
\caption{Estimation of QFM: Serially and Cross-Sectionally Correlated Idiosyncratic Errors}

\begin{tabular}{cc|cccc|cccc|cccc}
\hline
          &         &  \multicolumn{4}{|c|}{ $\tau=0.25$ }    &  \multicolumn{4}{c|}{ $\tau=0.5$ }    &  \multicolumn{4}{c}{ $\tau=0.75$ }  \\
  $N$ &  $T$ &   $\hat{r}$     &   $f_{1t}$   &      $f_{2t}$  &    $f_{3t}$   &  $\hat{r}$     &   $f_{1t}$   &      $f_{2t}$  &    $f_{3t}$ &  $\hat{r}$     &   $f_{1t}$   &      $f_{2t}$  &    $f_{3t}$     \\
\hline
 50&50& 2.54&0.926&0.705 &0.409 & 2.16& 0.952 & 0.808 &0.029  & 2.53 & 0.921 &0.700 &0.423 \\
 50&100& 2.49 &0.941 &0.703&0.397 & 1.95 &0.959 & 0.845 & 0.001 & 2.50 & 0.934 &0.723 &0.423\\
  50&200& 2.66 &0.945&0.803&0.460 &  1.97&0.963&0.881 &0.000 & 2.64 &0.939 &0.756 &0.471\\
\hline
 100&50&2.52 &0.942&0.780 &0.495 & 2.02& 0.977 &0.820 &0.021  &2.41& 0.946 &0.744 &0.472 \\
 100&100&2.91&0.976 &0.896&0.697 & 2.06&0.981 & 0.945 & 0.006 &2.87& 0.977 &0.893 &0.686\\
  100&200& 2.90&0.979&0.924&0.702 & 2.01&0.983 & 0.966 &0.000 & 2.92&0.980 &0.933 &0.713\\
  \hline
 200&50&2.47 &0.967&0.732 &0.569 & 2.05 & 0.987 &0.870 &0.032 & 2.52& 0.969 &0.785 &0.576 \\
200&100&2.88 &0.989&0.913 &0.802 & 2.00 & 0.991 &0.982 &0.000 & 2.89& 0.989 &0.938 &0.788 \\
  200&200& 3.00&0.990&0.982&0.866 &2.00 &0.992&0.983 &0.000 & 3.00&0.990 &0.981 &0.866\\
  \hline
\end{tabular}
\begin{tablenotes}
      \small
      \item Note: The DGP considered in this Table is: $X_{it} = \lambda_{1i} f_{1t} +  \lambda_{2i} f_{2t} + ( \lambda_{3i} f_{3t}) \cdot e_{it}$, $f_{1t} =0.8 f_{1,t-1}+\epsilon_{1t}$, $f_{2t} =0.5 f_{2,t-1}+\epsilon_{2t}$, $f_{3t} =|g_t|$, $\lambda_{1i},\lambda_{2i},\epsilon_{1t},\epsilon_{2t},g_t \sim i.i.d \text{ }\mathcal{N}(0,1)$, and $\lambda_{3i} \sim i.i.d \text{ }U[1,2]$. $e_{it} = \beta e_{i,t-1} + v_{it} + \rho \cdot \sum_{j=i-J,j\neq i}^{i+J}v_{jt}$, $v_{it}\sim i.i.d \text{ }\mathcal{N}(0,1)$, $\beta=\rho=0.2$ and $J=3$. For each $\tau$, the first column reports the average rank estimator $\hat{r}$ from 1000 replications, while the second to fourth columns report the average $R^2$ in the regression of (each of) the true factors on the QFA factors $\hat{F}^{\tau}_{QFA}$, obtained from the IQR algorithm.
    \end{tablenotes}
\end{threeparttable}
\end{center}
\end{table}

\begin{table}[htp]
\begin{center}
\begin{threeparttable}
\caption{All Empirical Applications: Number of Factors}

\begin{tabular}{c ccc}
\hline
& \textbf{Climate} &\textbf{Macro} & \textbf{Finance}\\
$(N,T)$ & (441,102)& (211,238)  & (429,420) \\
\hline
$PC_{p1}$ & 8 & 8 & 4 \\
ER & 2&1&1 \\
\hline
$\hat{r}_{\text{rank}}$ $\tau=0.01$ &1&1&1 \\
$\hat{r}_{\text{rank}}$ $\tau=0.05$&1&1&1\\
$\hat{r}_{\text{rank}}$ $\tau=0.10$&2&2&1\\
$\hat{r}_{\text{rank}}$ $\tau=0.25$ &4&4&1\\
$\hat{r}_{\text{rank}}$ $\tau=0.50$ &6&5&2 \\
$\hat{r}_{\text{rank}}$ $\tau=0.75$ &4&5&1 \\
$\hat{r}_{\text{rank}}$ $\tau=0.90$ &2&2&1 \\
$\hat{r}_{\text{rank}}$ $\tau=0.95$ &1&1&1 \\
$\hat{r}_{\text{rank}}$ $\tau=0.99$ &1&1&1 \\
\hline
\end{tabular}
\begin{tablenotes}
      \small
      \item Note: This Table provides the estimated numbers of mean factors using $PC_{p1}$ of \textcolor{blue}{BN (2002)}, the ER estimator of \citet{ahn2013eigenvalue}, and the estimated numbers of quantile factors at $\tau\in\{0.01, 0.05,0.1,0.25,0.75,0.9,0.95,0.99\}$ using the rank-minimization estimator proposed in subsection 3.2.    \end{tablenotes}
\end{threeparttable}
\end{center}
\end{table}

\begin{table}[htp]
\begin{center}
\begin{threeparttable}
\caption{All Empirical Applications: Comparison of $\hat{F}_{QFA}$ and $\hat{F}_{PCA}$}
\begin{tabular}{cc|cccccc}
\hline
& &   \multicolumn{6}{c}{Elements of $\hat{F}_{QFA}^{\tau}$} \\
 Dataset &$\tau$& 1 & 2 & 3 & 4 & 5 & 6 \\
\hline
\textbf{Climate} & $0.01$ & 0.599&&& &&\\
& $0.05$ &0.623&& &&&\\
& $0.10$ &0.759&0.848&&&&\\
& $0.25$ & 0.939&0.961& 0.965&0.941& &\\
& $0.50$ &0.995&0.995&0.992&0.988&0.980&0.970\\
& $0.75$ & 0.950&0.961&0.966&0.933 &&\\
& $0.90$ &0.755&0.905&&&&\\
& $0.95$ &0.629&&&&&\\
& $0.99$ & 0.567&&&&&\\
 \hline
\textbf{Macro} & $0.01$ & 0.657&&& &&\\
& $0.05$ &0.733&& &&&\\
& $0.10$ &0.796&0.871&&&&\\
& $0.25$ & 0.952&0.932& 0.939&0.890&&\\
& $0.50$ &0.993&0.976&0.964&0.945&0.923&\\
& $0.75$ & 0.906&0.945&0.943&0.903 &0.882&\\
& $0.90$ &0.316&0.911&&&&\\
& $0.95$ &0.261&&&&&\\
& $0.99$ & 0.266&&&&&\\
 \hline
 \textbf{Finance} & $0.01$ & 0.560&&& &&\\
& $0.05$ &0.731&& &&&\\
& $0.10$ &0.803&&&&&\\
& $0.25$ & 0.921&& &&&\\
& $0.50$ &0.993&0.977&&&&\\
& $0.75$ & 0.945&&& &&\\
& $0.90$ &0.783&&&&&\\
& $0.95$ &0.660&&&&&\\
& $0.99$ & 0.492&&&&&\\
 \hline
\end{tabular}
\begin{tablenotes}
      \small
      \item Note: This Table reports the $R^2$ of regressing each element of $\hat{F}_{QFA}$ on $\hat{F}_{PCA}$. For $\hat{F}_{QFA}$, the numbers of estimated factors is obtained from Table 7 while, for $\hat{F}_{PCA}$, the numbers of estimated factors are 8 for all datasets.     \end{tablenotes}
\end{threeparttable}
\end{center}
\end{table}

\begin{table}[htp]
\begin{center}
\begin{threeparttable}
\caption{Climate: P-values of Granger Non-Causality Tests}
\begin{tabular}{c|cccccc}
\hline
  &   \multicolumn{6}{c}{Elements of $\hat{F}_{QFA}^{\tau}$} \\
$\tau$ & 1 & 2 & 3 & 4 & 5 & 6 \\
\hline
 $0.01$ & \textbf{0.004}&&& &&\\
$0.05$ &\textbf{0.010}&& &&&\\
$0.10$ &0.225&0.336&&&&\\
$0.25$ & 0.231&0.371& 0.093&0.834&&\\
$0.50$ &0.462&0.457&\textbf{0.027}&0.362&0.808&\textbf{0.037}\\
$0.75$ & 0.381&0.404&0.229&0.423 &&\\
$0.90$ &0.340&0.769&&&&\\
 $0.95$ &0.621&&&&&\\
 $0.99$ & 0.958&&&&&\\
 \hline
\end{tabular}
\begin{tablenotes}
      \small
      \item Note: This Table reports the p-values of Granger non-causality tests, where each of the QFA factors is regressed on their own lags and the lags of $\Delta \log ( \text{CO}_2)$, and the lag lengths are chosen according to BIC.   \end{tablenotes}
\end{threeparttable}
\end{center}
\end{table}

\begin{table}[htp]
\begin{center}
\begin{threeparttable}
\caption{Macro Forecasting: MSE of Different Methods}
\begin{tabular}{lcccccc}
\hline
&\multicolumn{2}{c}{Pre-Crisis}&\multicolumn{2}{c}{Crisis- Post-Crisis}&\multicolumn{2}{c}{Full}\\
 &MSE& R. MSE& MSE& R. MSE&MSE& R. MSE\\
 \hline
 &\multicolumn{6}{c}{\textit{Real GDP Growth}} \\
AR Benchmark&4.526&1.000&5.456&1.000&4.904&1.000\\
AR + $\hat{F}_{PCA}$ &4.282&0.946&5.373&0.985&4.725&0.964\\
AR + $\hat{F}_{PCA}$ + $\hat{F}_{QFA}^{90}$ &4.155&\textbf{0.918}&5.331&0.977&4.634&\textbf{0.945}\\
AR + $\hat{F}_{PCA}$ + $\hat{F}_{QFA}^{99}$ &4.354&0.962&5.270&\textbf{0.966}&4.728&0.964\\
AR + $\hat{F}_{PCA}$ + $\hat{F}_{QFA}^{90}$ + $\hat{F}_{QFA}^{99}$ &4.191&0.926&5.456&1.000&4.688&0.956\\
\hline
&\multicolumn{6}{c}{\textit{Inflation}} \\
AR Benchmark&0.266&1.000&0.790&1.000&0.479&1.000\\
AR + $\hat{F}_{PCA}$ &0.246&0.926&0.732&0.926&0.444&0.926\\
AR + $\hat{F}_{PCA}$ + $\hat{F}_{QFA}^{90}$ &0.246&0.926&0.732&0.926&0.444&0.927\\
AR + $\hat{F}_{PCA}$ + $\hat{F}_{QFA}^{99}$ &0.247&0.927&0.739&0.935&0.447&0.932\\
AR + $\hat{F}_{PCA}$ + $\hat{F}_{QFA}^{90}$ + $\hat{F}_{QFA}^{99}$ &0.245&\textbf{0.922}&0.726&\textbf{0.919}&0.441&\textbf{0.920}\\
  \hline
\end{tabular}
\begin{tablenotes}
      \small
      \item Note: This Table reports the MSE of five alternative 1-quarter-ahead forecasting methods for real GDP growth and inflation, and their relative MSE (R. MSE) compared with the AR benchmark model (the lowest R. MSE are shown in bold characters). The out-of-sample forecasting is implemented using rolling windows with 120 observations. The full forecasting evaluation period is from 1990Q1 to 2019Q2, the pre-crisis period is from 1990Q1 to 2007Q2, and the crisis plus post-crisis period is from 2007Q3 to 2019Q2.  \end{tablenotes}
\end{threeparttable}
\end{center}
\end{table}

\newpage
\clearpage
\begin{figure}[H]
\caption{Normal Approximations of the Estimated Factors using SQR for $N=T=50$.}
\subfloat{\includegraphics[width = 3in]{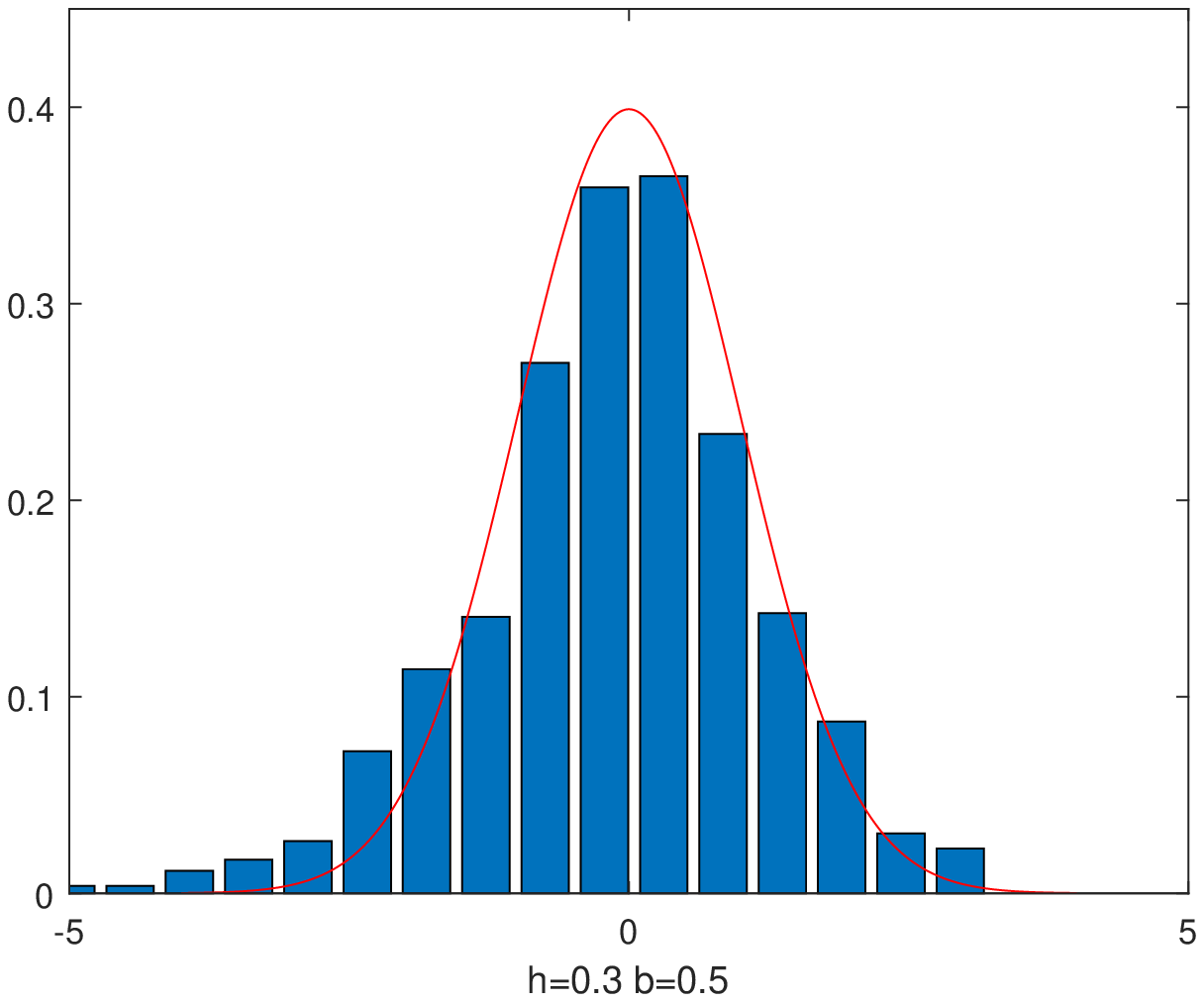}} 
\subfloat{\includegraphics[width = 3in]{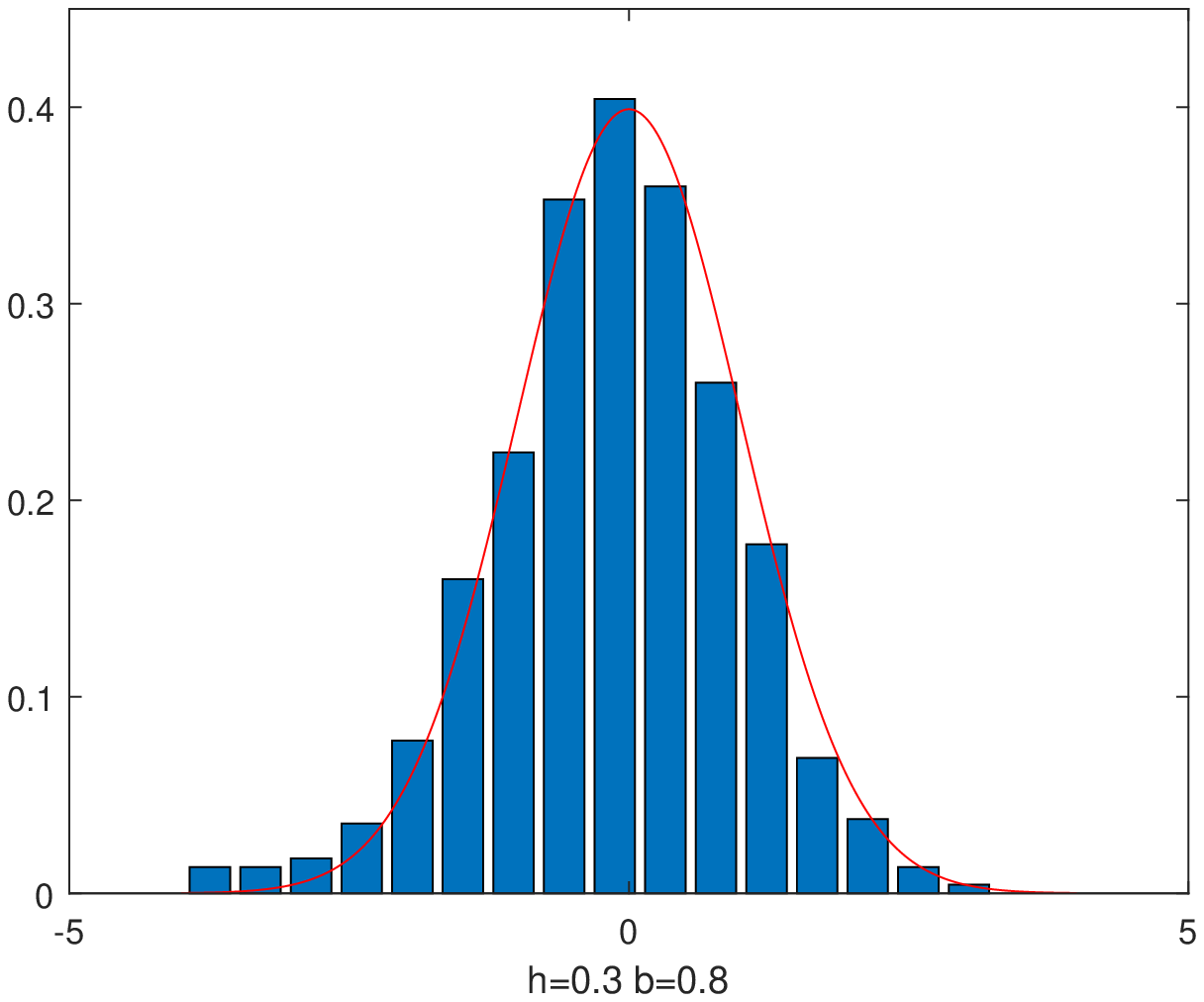}}
\\
\subfloat{\includegraphics[width = 3in]{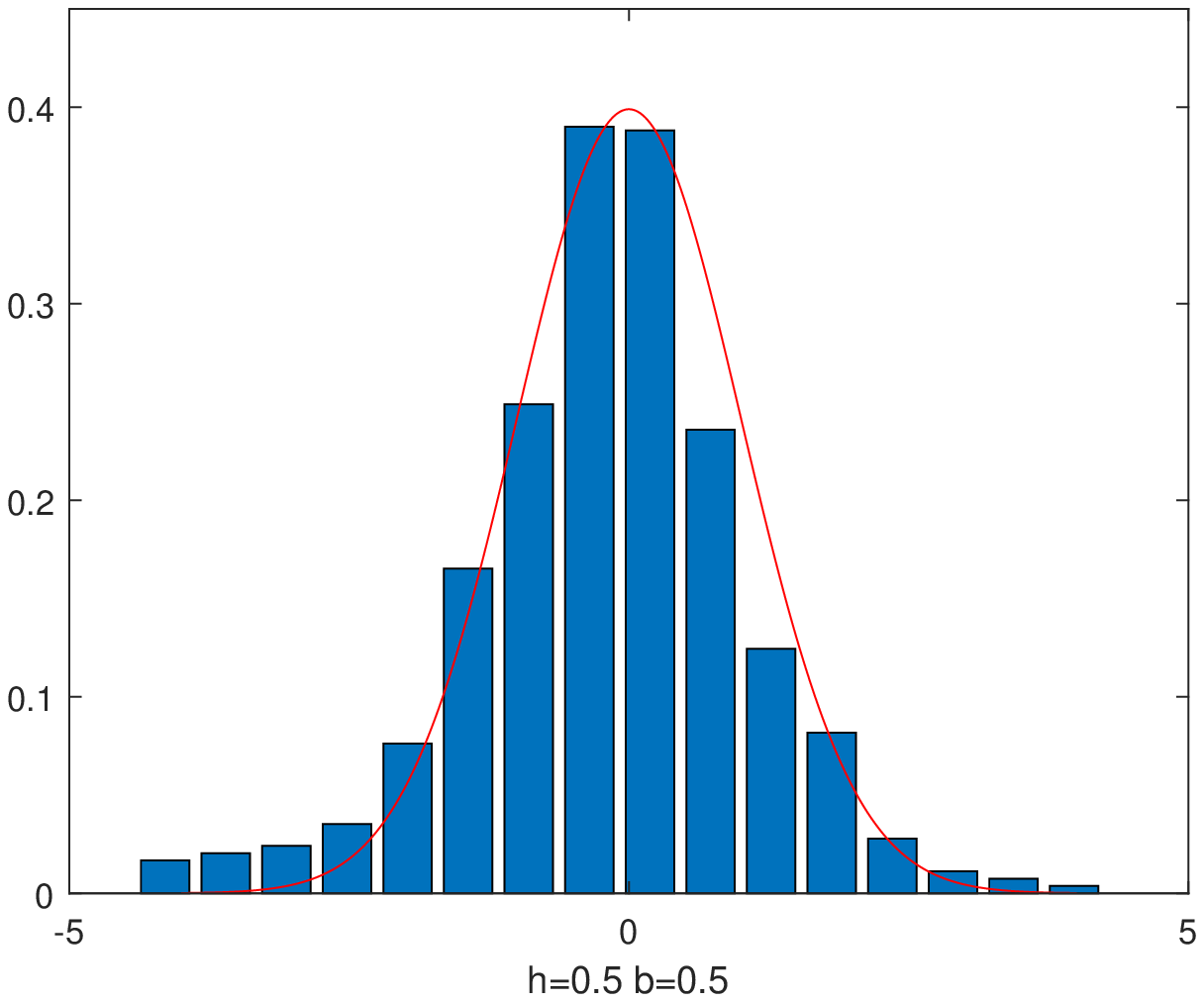}}
\subfloat{\includegraphics[width = 3in]{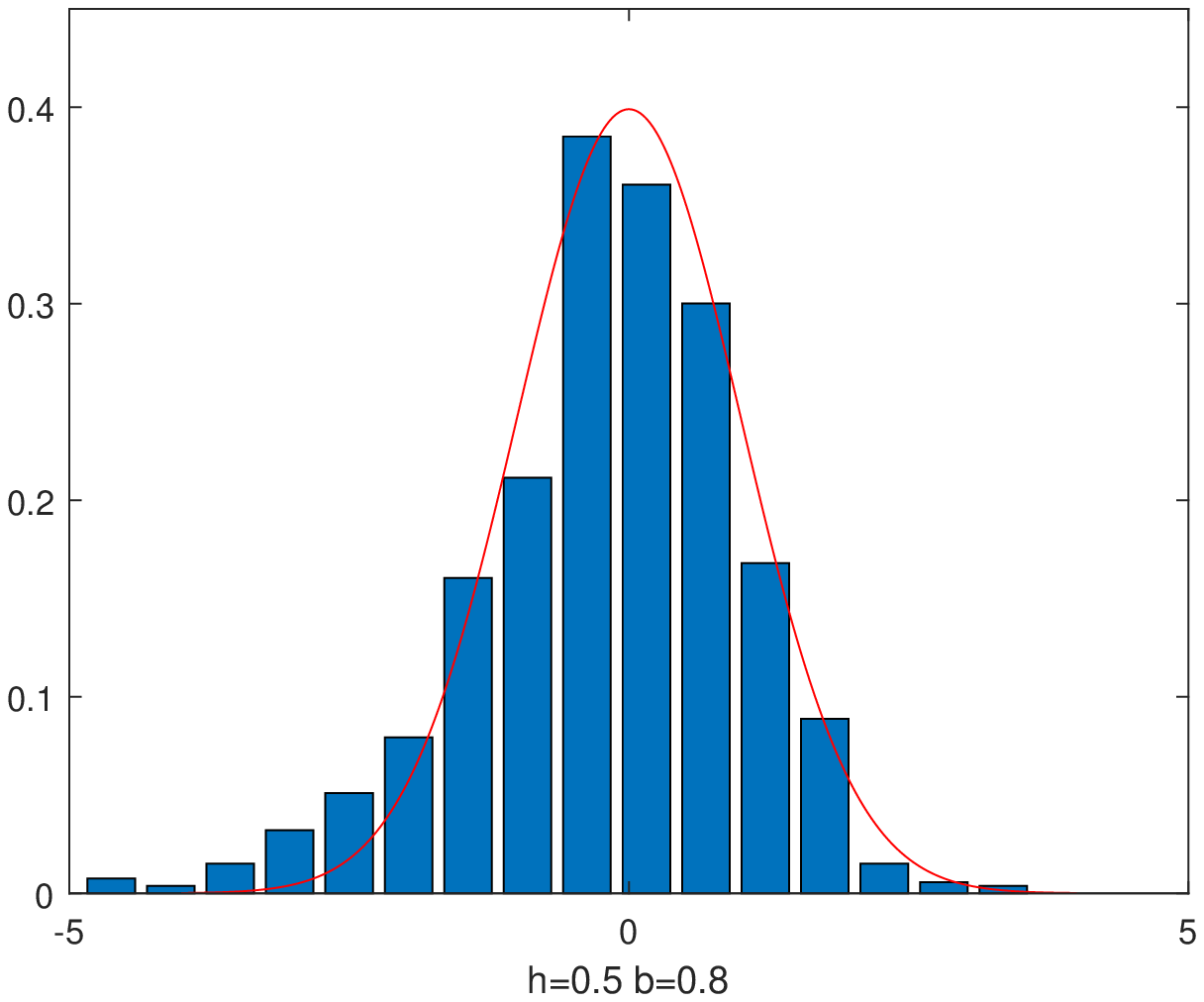}} 
\\
\subfloat{\includegraphics[width = 3in]{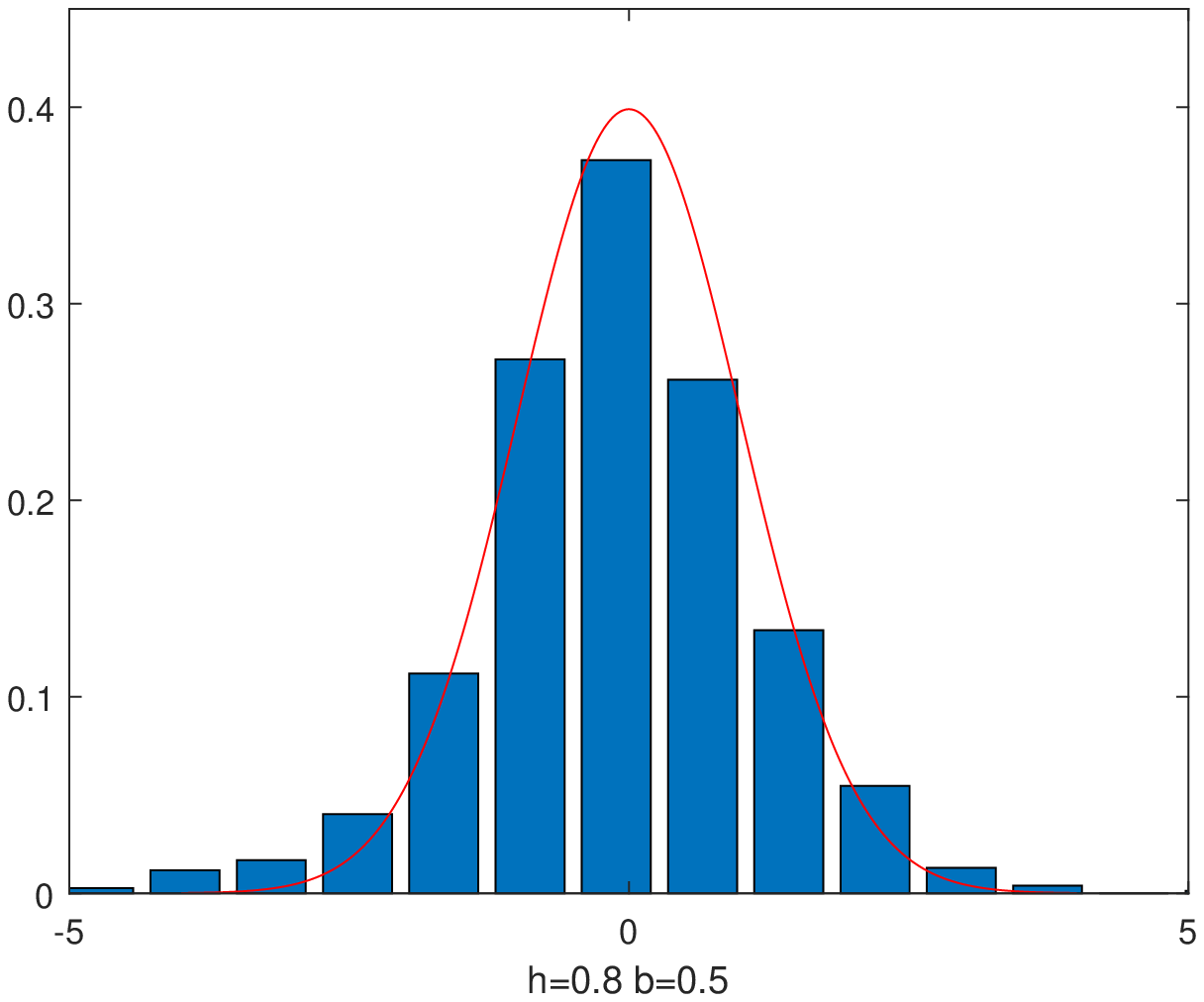}}
\subfloat{\includegraphics[width = 3in]{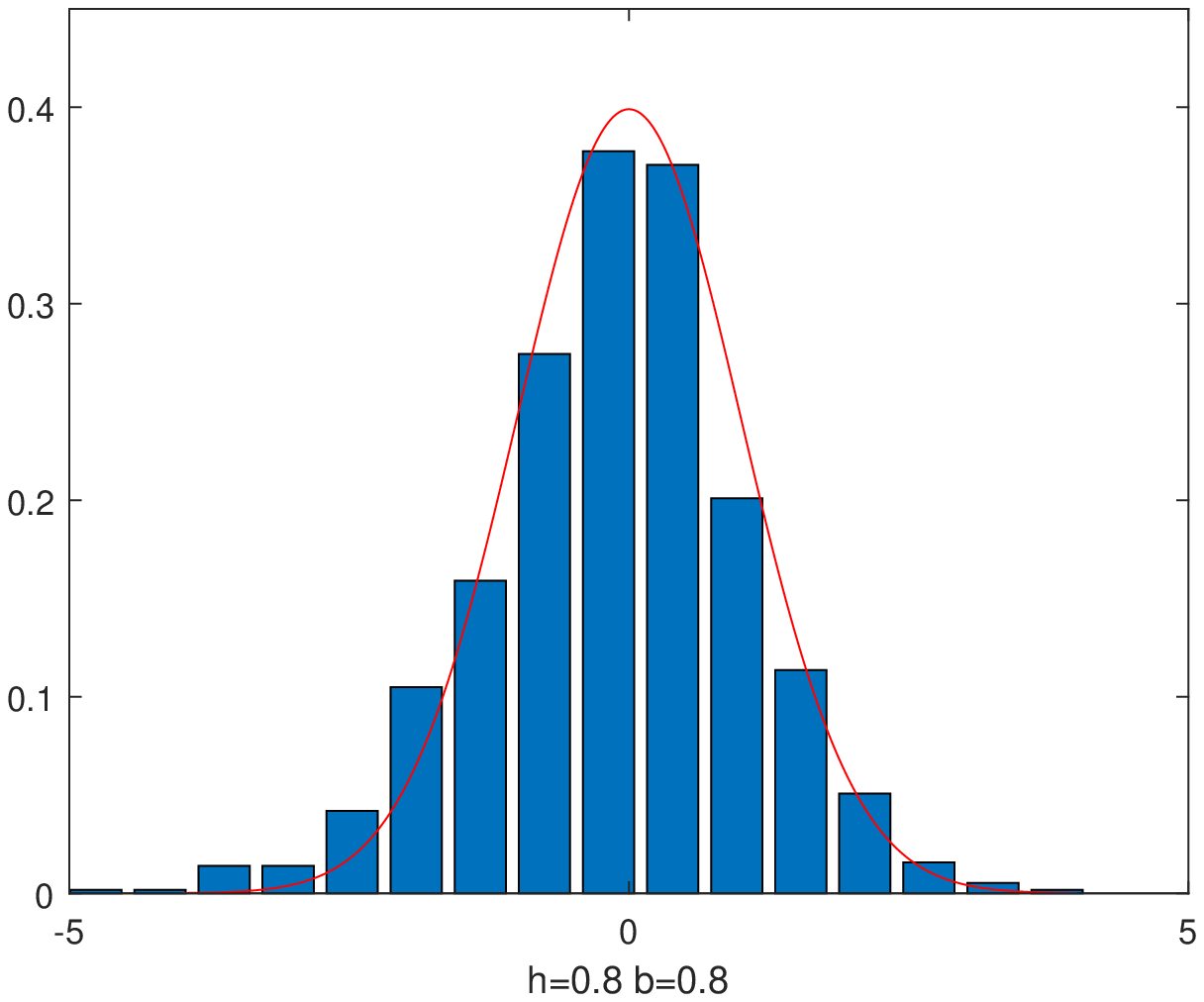}} 
\label{some example}
{\centering
\begin{minipage}{6.5in}
    \small
    DGP: $X_{it} = \lambda_i f_t + f_t \epsilon_{it}$, where $f_t \sim i.i.d \text{ }\mathcal{U}(1,2)$ and they are normalized such that $F'F/T=1$, $\lambda_{i} \sim i.i.d \text{ }\mathcal{N}(0,1)$ and $\epsilon_{it} \sim i.i.d \text{ }\mathcal{N}(0,1)$. The figure plots the histograms of the standardized estimators of the factors using SQR: $\hat{V}_{f_t}^{-1}\sqrt{N}(\tilde{f}_t -f_{0t})$ at $\tau=0.25, t=T/2$ from 1000 replications, where $\hat{V}_{f_t}$ is estimated using the formula in Remark 4.4, $h$ is the bandwidth parameter in the smoothed check function, and $b$ is the bandwidth parameter used in $\hat{V}_{f_t}$.
    \end{minipage}
    }
\end{figure}

\newpage
\clearpage
\begin{figure}[H]
\caption{Normal Approximations of the Estimated Factors using SQR for $N=T=200$.}
\subfloat{\includegraphics[width = 3in]{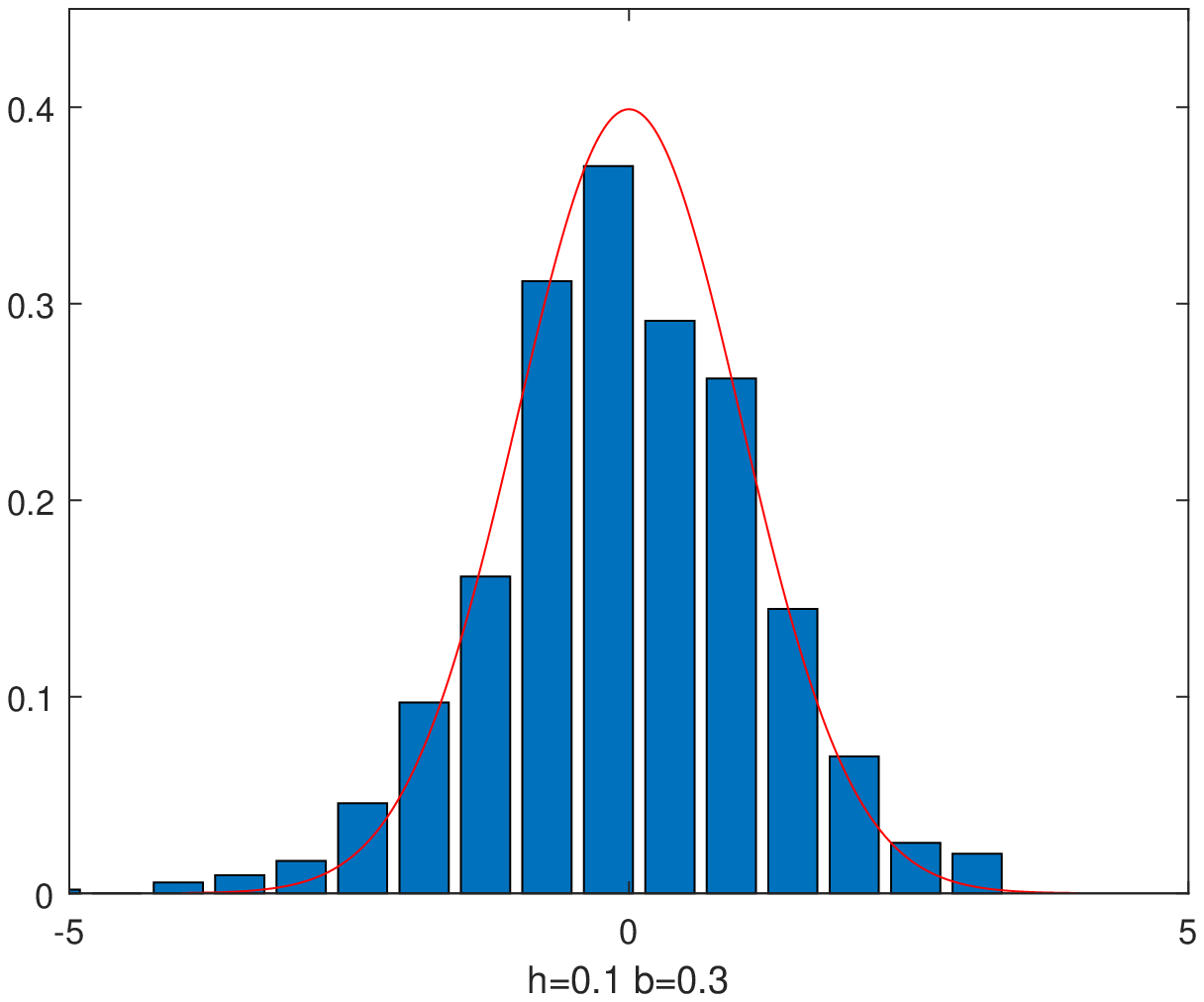}} 
\subfloat{\includegraphics[width = 3in]{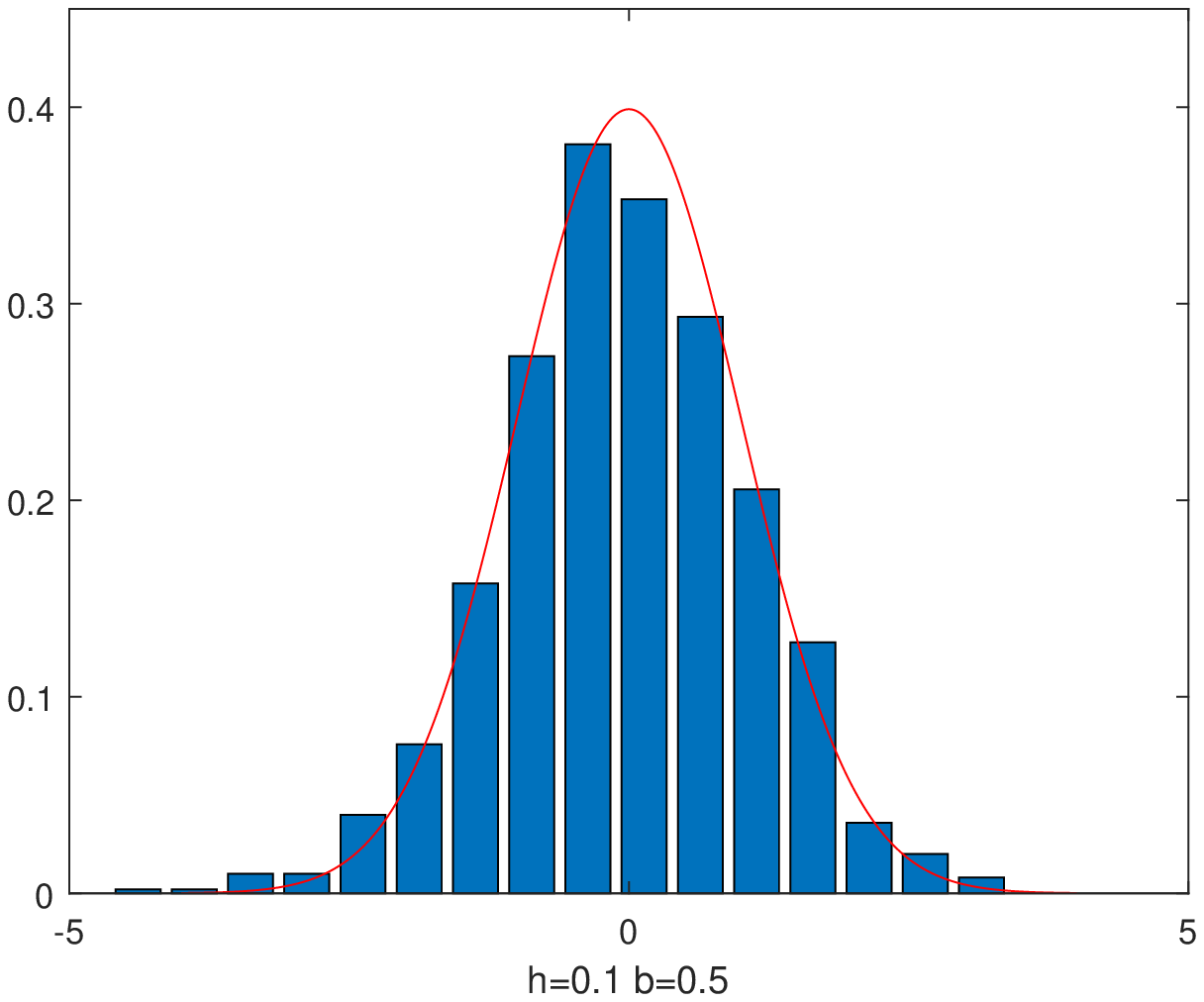}}
\\
\subfloat{\includegraphics[width = 3in]{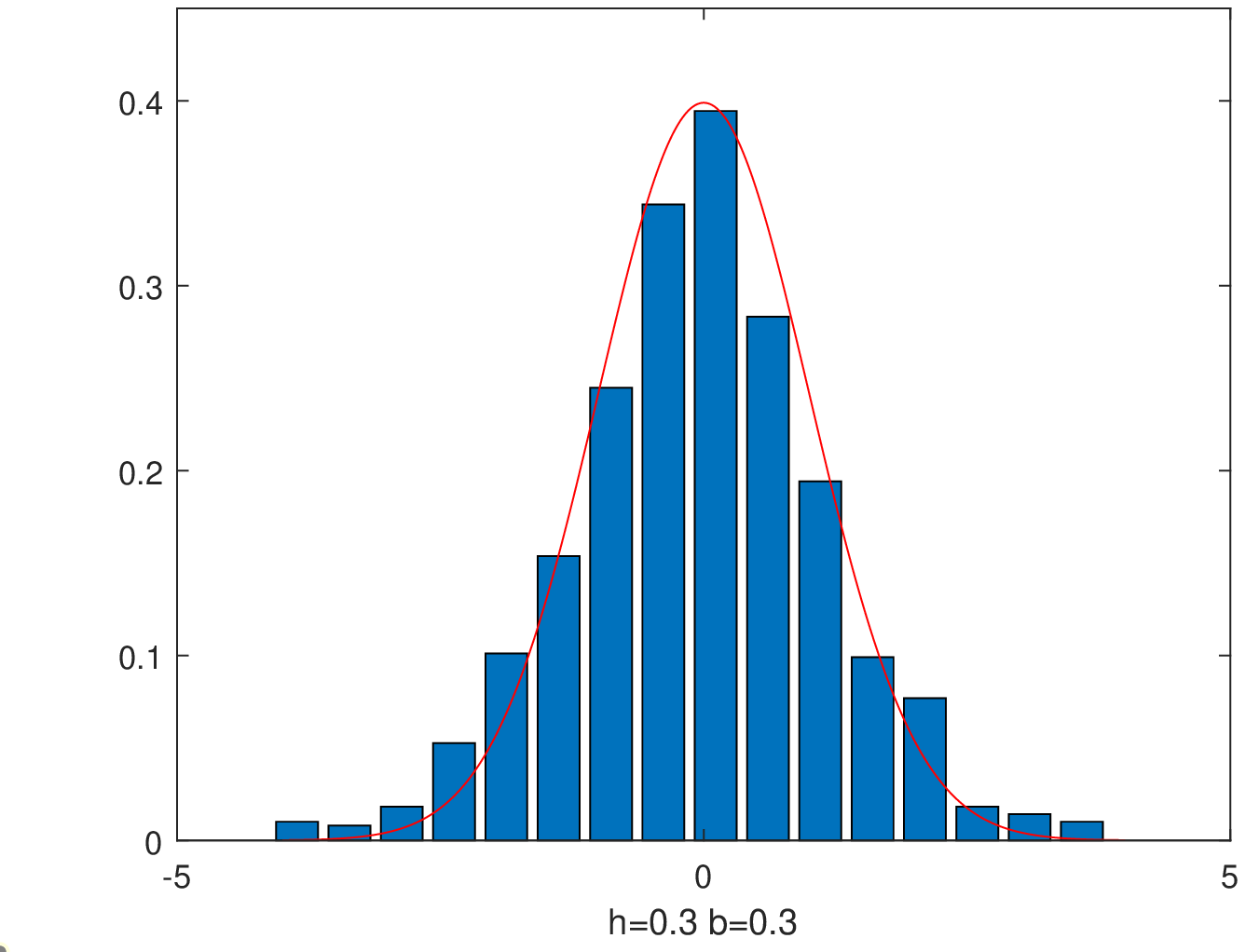}}
\subfloat{\includegraphics[width = 3in]{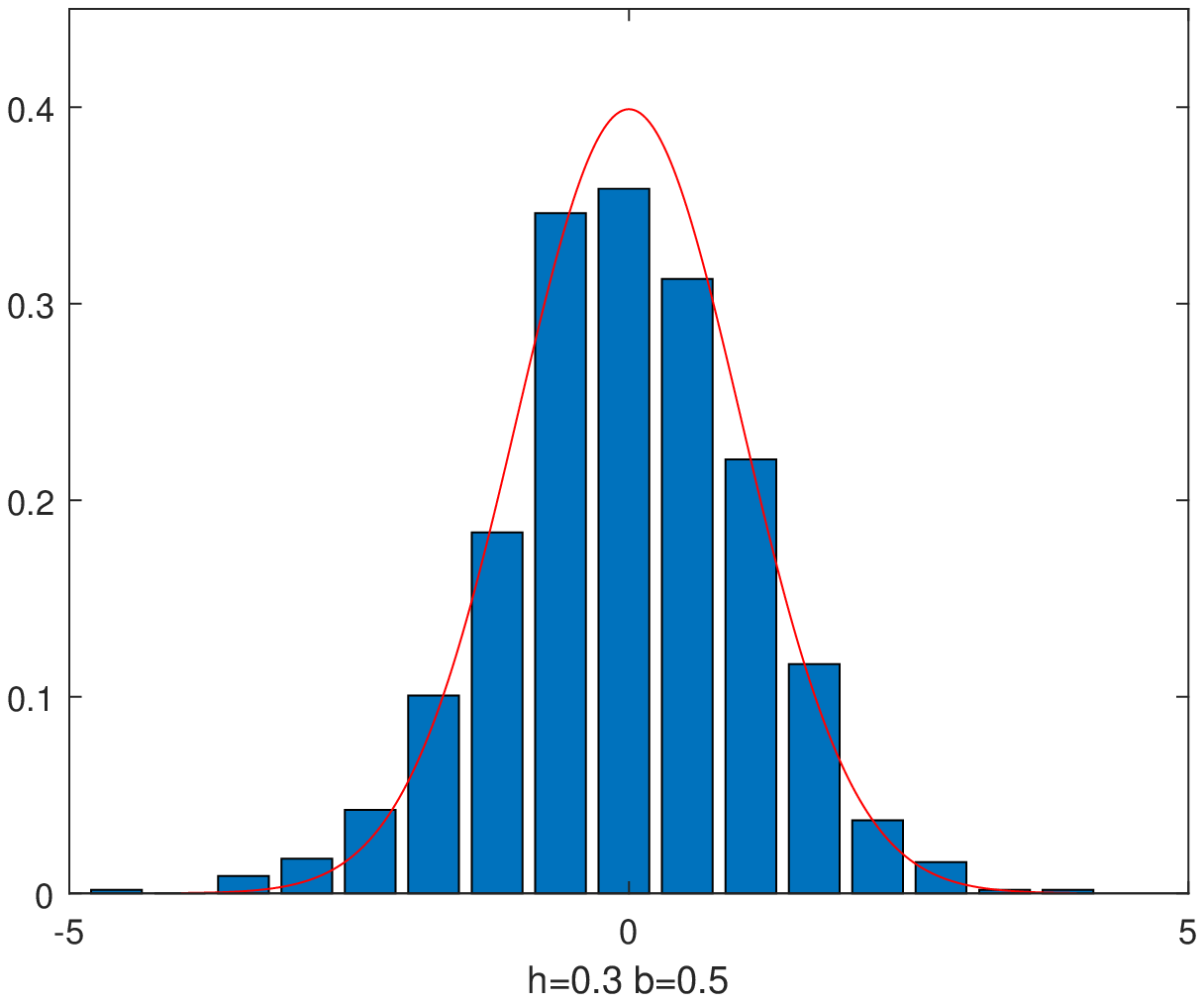}} 
\\
\subfloat{\includegraphics[width = 3in]{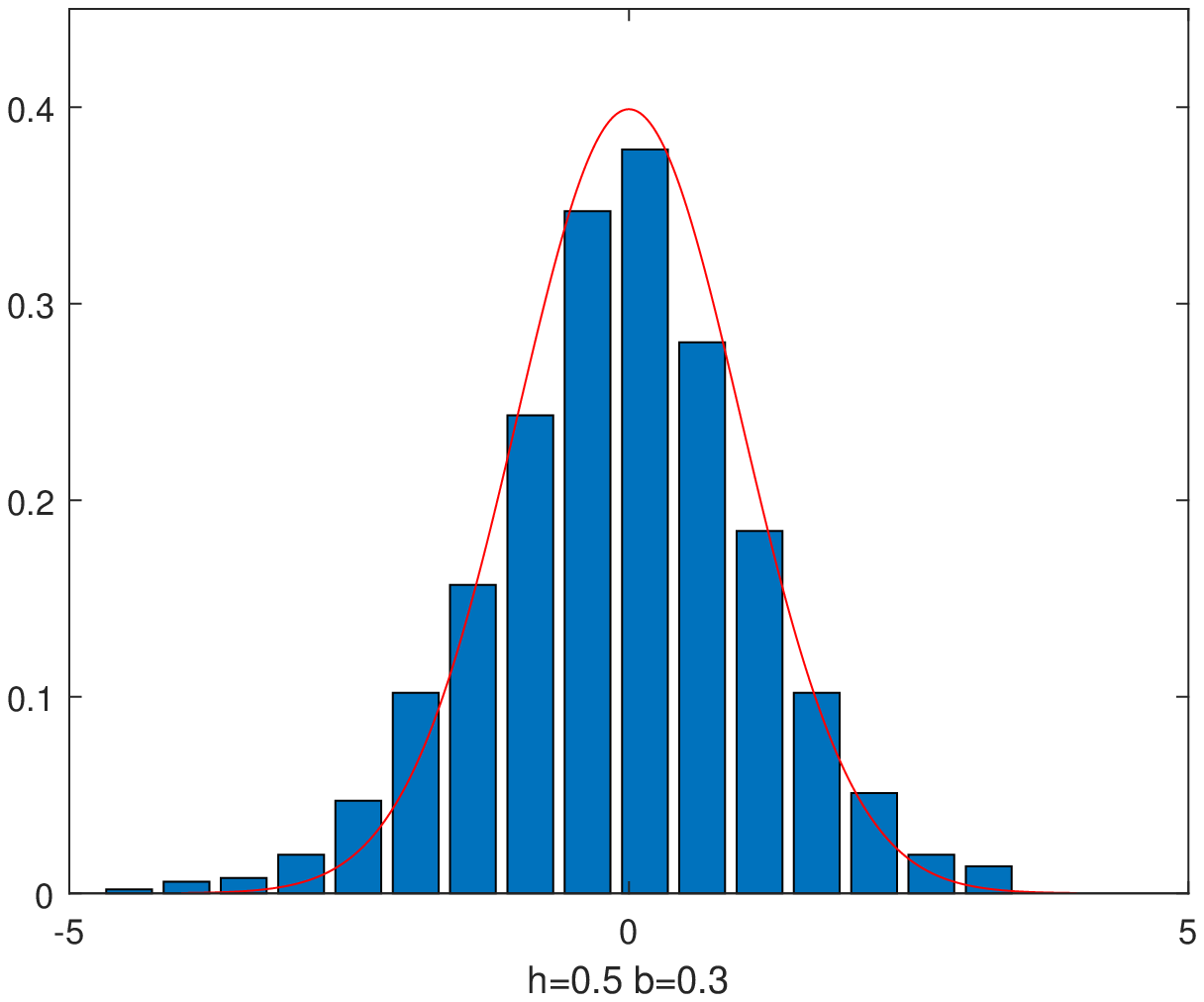}}
\subfloat{\includegraphics[width = 3in]{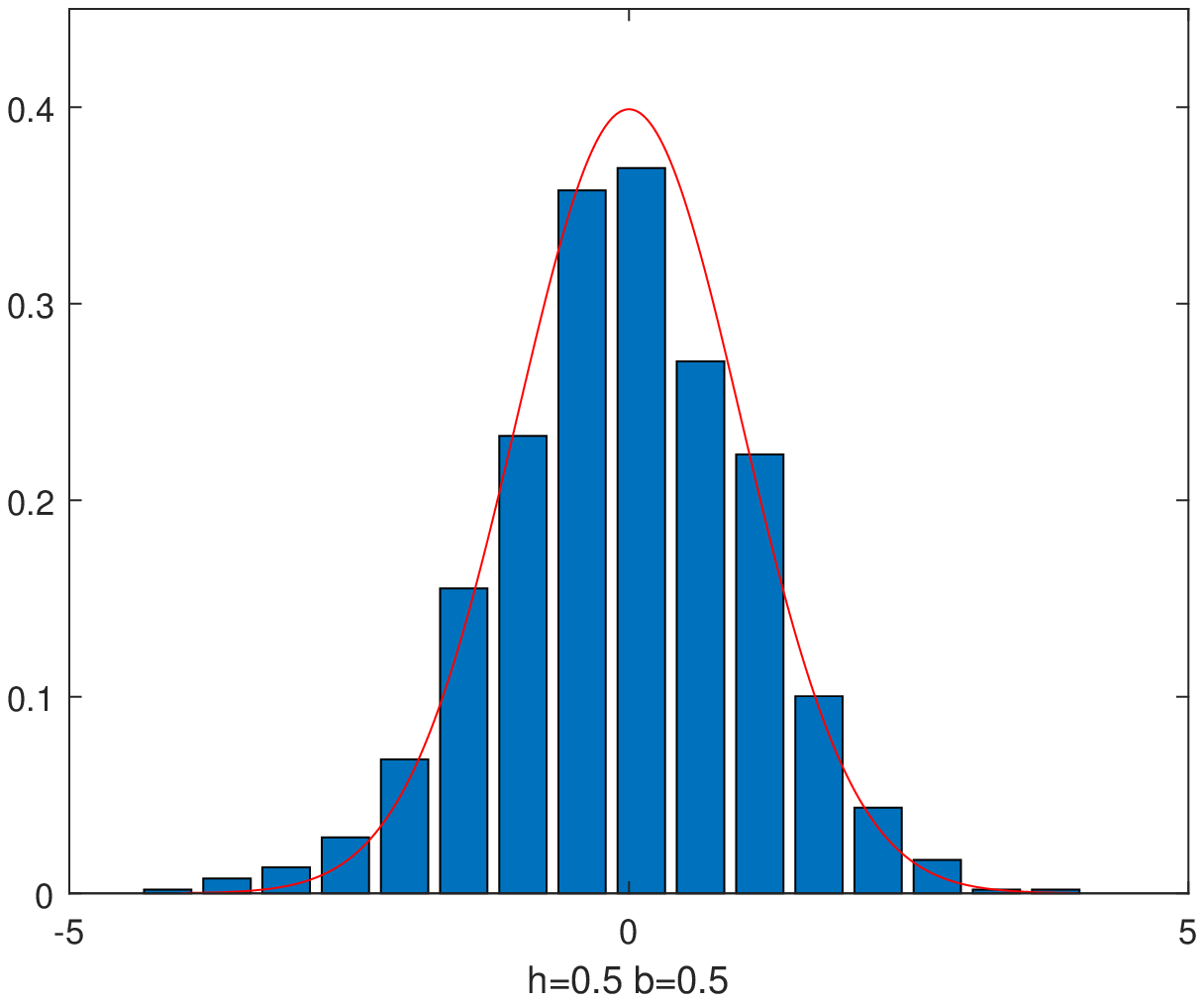}} 
\label{some example}
{\centering
\begin{minipage}{6.5in}
    \small
    DGP: $X_{it} = \lambda_i f_t + f_t \epsilon_{it}$, where $f_t \sim i.i.d \text{ }\mathcal{U}(1,2)$ and they are normalized such that $F'F/T=1$, $\lambda_{i} \sim i.i.d \text{ }\mathcal{N}(0,1)$ and $\epsilon_{it} \sim i.i.d \text{ }\mathcal{N}(0,1)$. The figure plots the histograms of the standardized estimators of the factors using SQR: $\hat{V}_{f_t}^{-1}\sqrt{N}(\tilde{f}_t -f_{0t})$ at $\tau=0.25, t=T/2$ from 1000 replications, where $\hat{V}_{f_t}$ is estimated using the formula in Remark 4.4, $h$ is the bandwidth parameter in the smoothed check function, and $b$ is the bandwidth parameter used in $\hat{V}_{f_t}$.
    \end{minipage}
    }
\end{figure}

\newpage
\clearpage
{\centering
\begin{figure}[H]
\caption{Macro Forecasting: Predictive Scores of Density Forecasts for GDP Growth and Inflation}
\subfloat{\includegraphics[width = 3.5in]{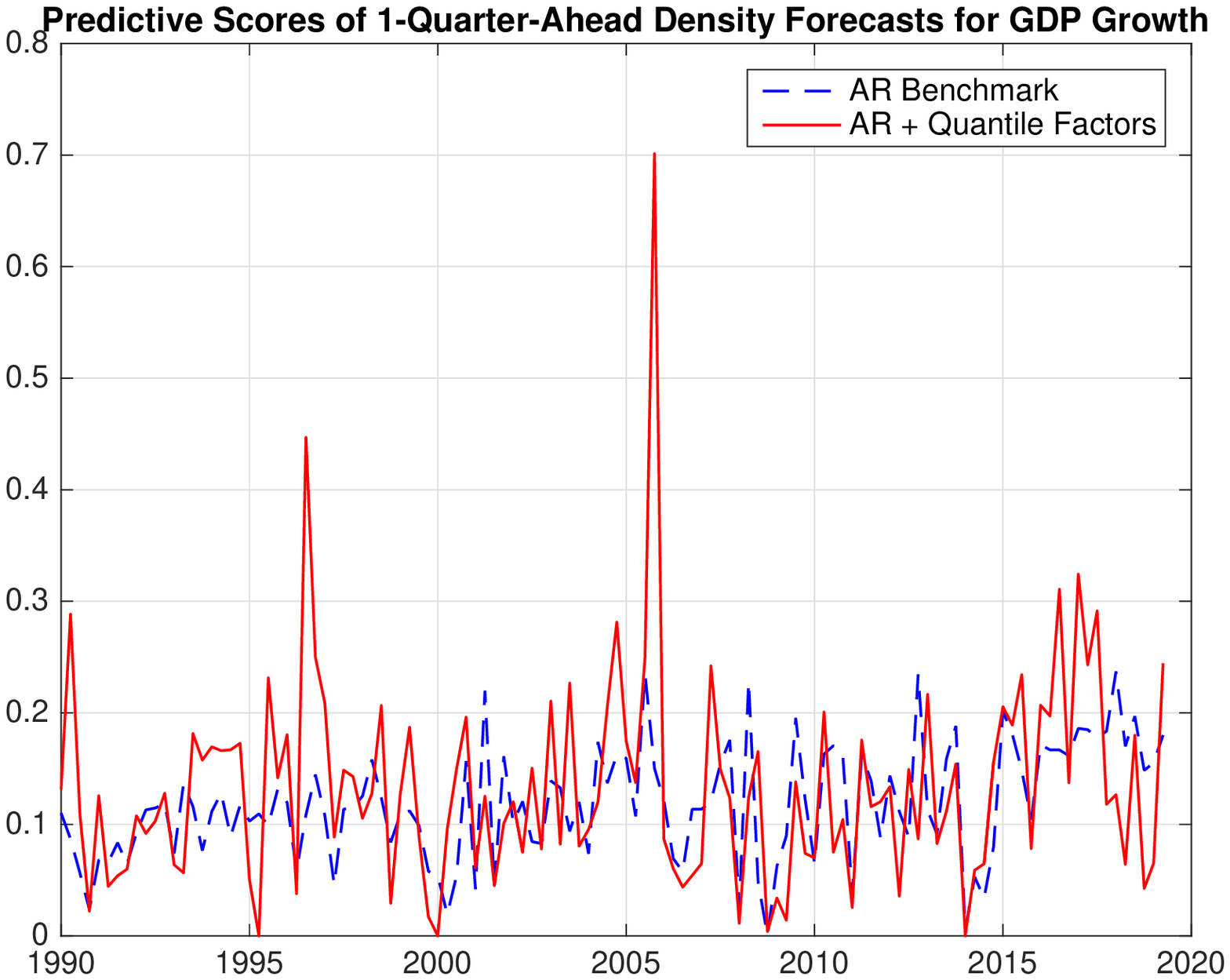}} 
\subfloat{\includegraphics[width = 3.5in]{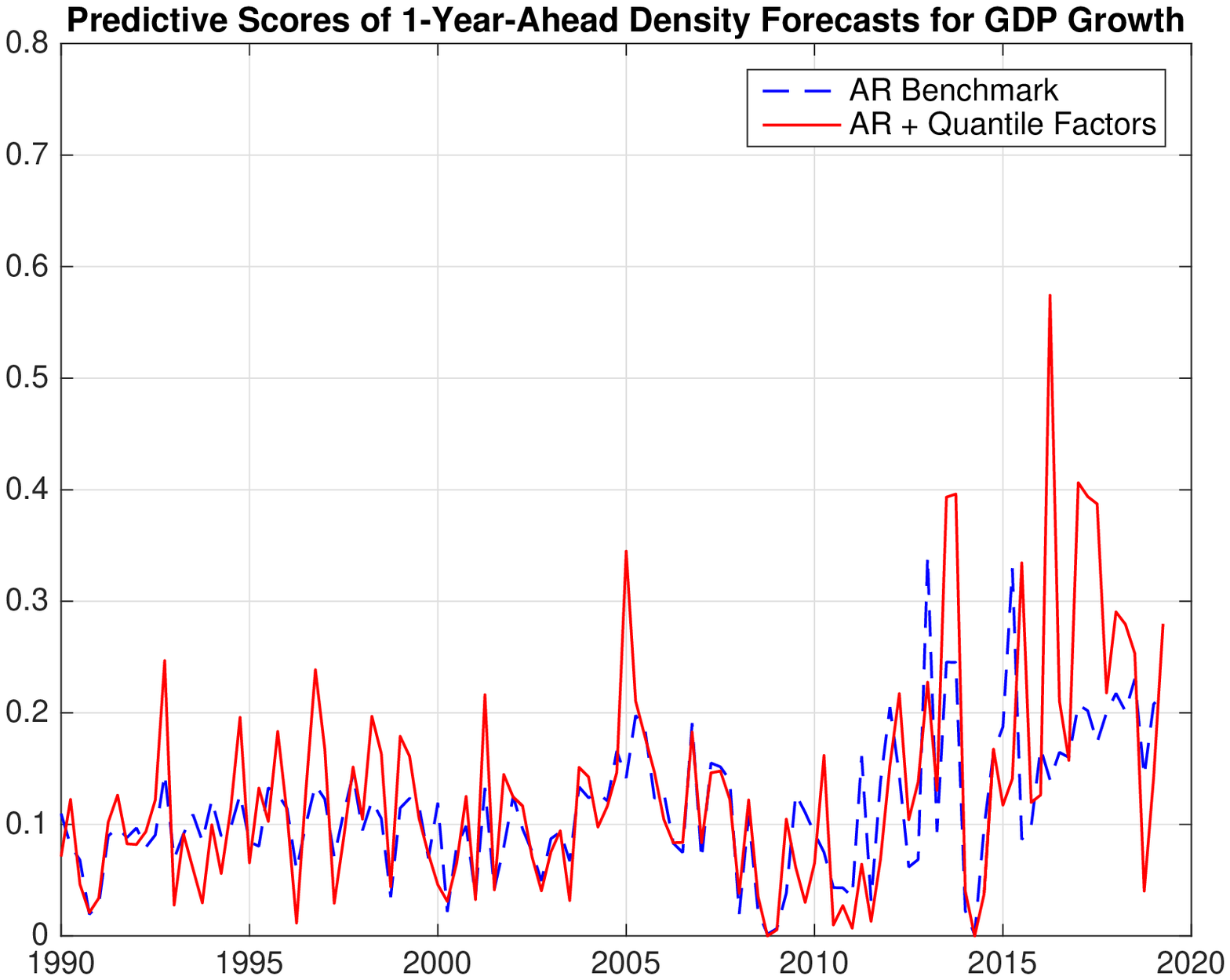}}
\\
\subfloat{\includegraphics[width = 3.5in]{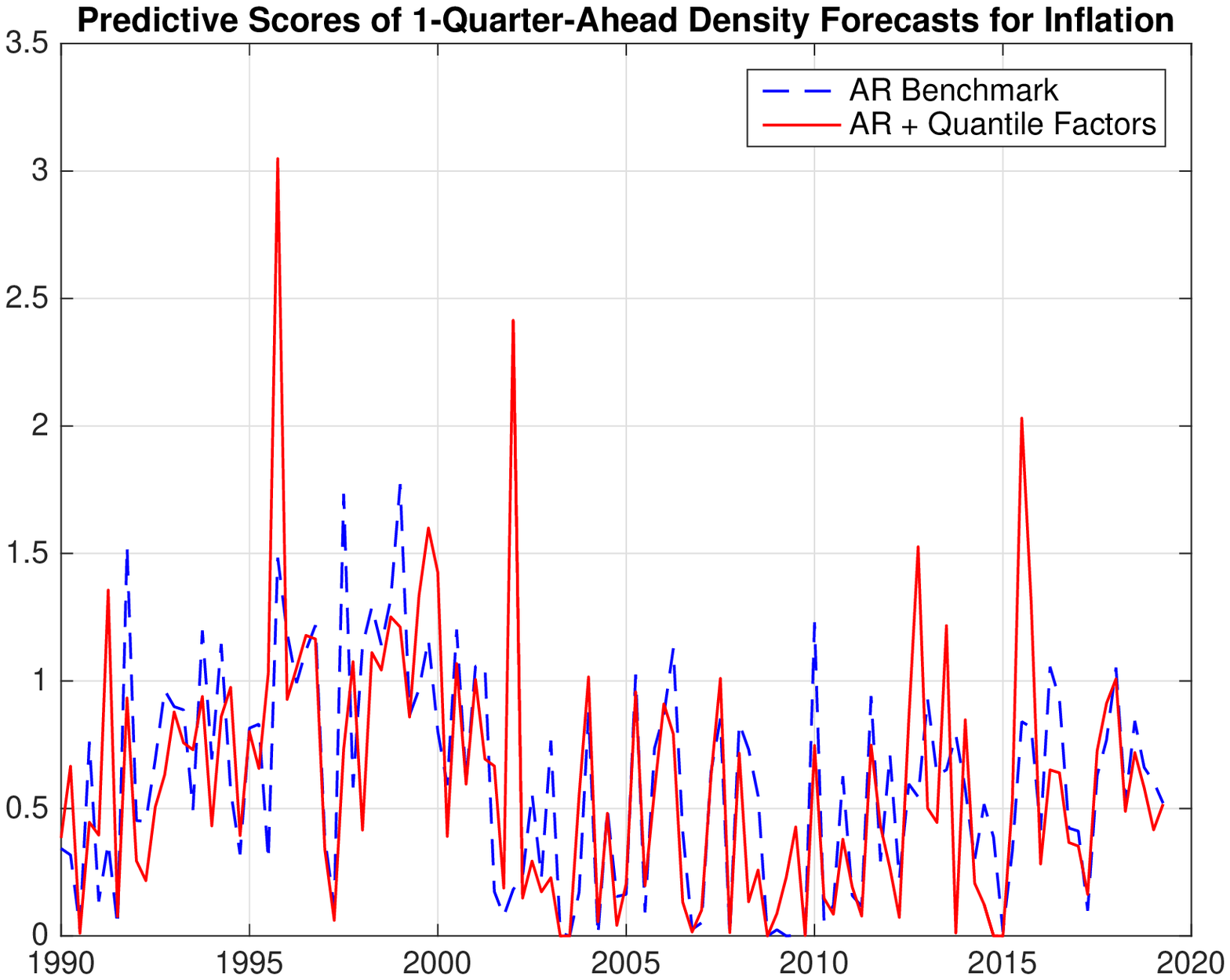}}
\subfloat{\includegraphics[width = 3.5in]{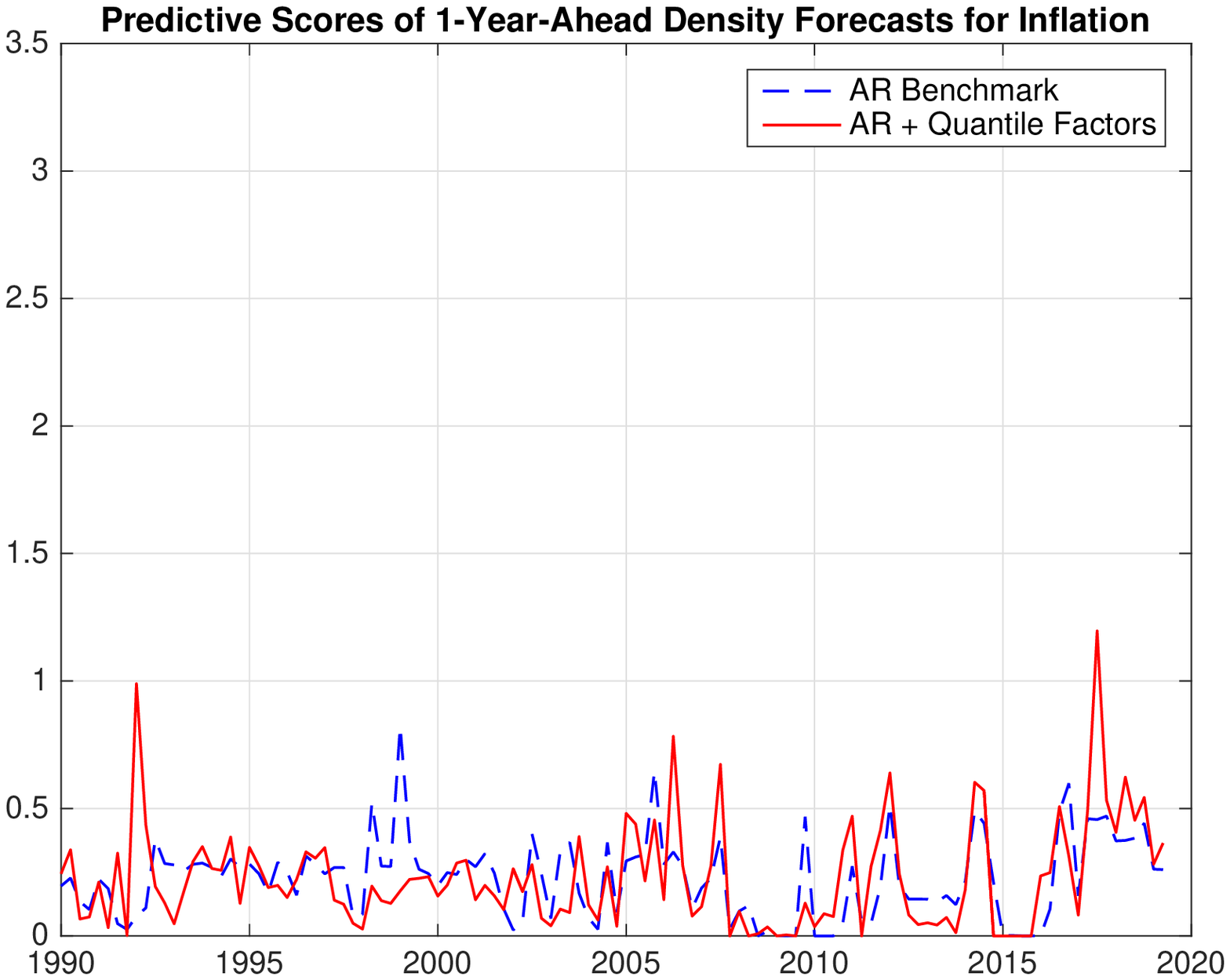}} 
\label{some example}
{\centering
\begin{minipage}{6.5in}
    \small
    Note: The graphs plot the predictive scores of 1-quarter-ahead and 1-year-ahead density forecasts for real GDP growth and inflation. The evaluation period is from 1990Q1 to 2019Q2, and the out-of-sample forecasting is implemented using rolling windows with 120 observations. The predicted $\tau$-quantiles are constructed using quantile regressions of the target variable on its owns lags and the estimated quantile factors at $\tau$ (denoted as $\hat{F}_{QFA}^{\tau}$). The predicted densities are constructed as the density functions of skewed t-distributions by matching the predicted quantiles of the target variable at $\tau\in\{0.05,0.25,0.75,0.95\}$. The predictive scores are the predicted densities evaluated at the realized values of the target variable. Higher scores indicate more accurate forecasts. The dotted blue line is the predictive scores of the benchmark AR model where only the lags of the target variable are used to predict the $\tau$-quantiles, the red line is the predictive scores of the model where $\hat{F}_{QFA}^{\tau}$ is also used to predict the $\tau$-quantiles.
    \end{minipage}
    }
\end{figure}
\par
}

\newpage
\clearpage
{\centering
\begin{figure}[H]
\caption{Finance: Quantile Factors, Interquantile Range and Volatility Factor}
\subfloat{\includegraphics[width = 6in]{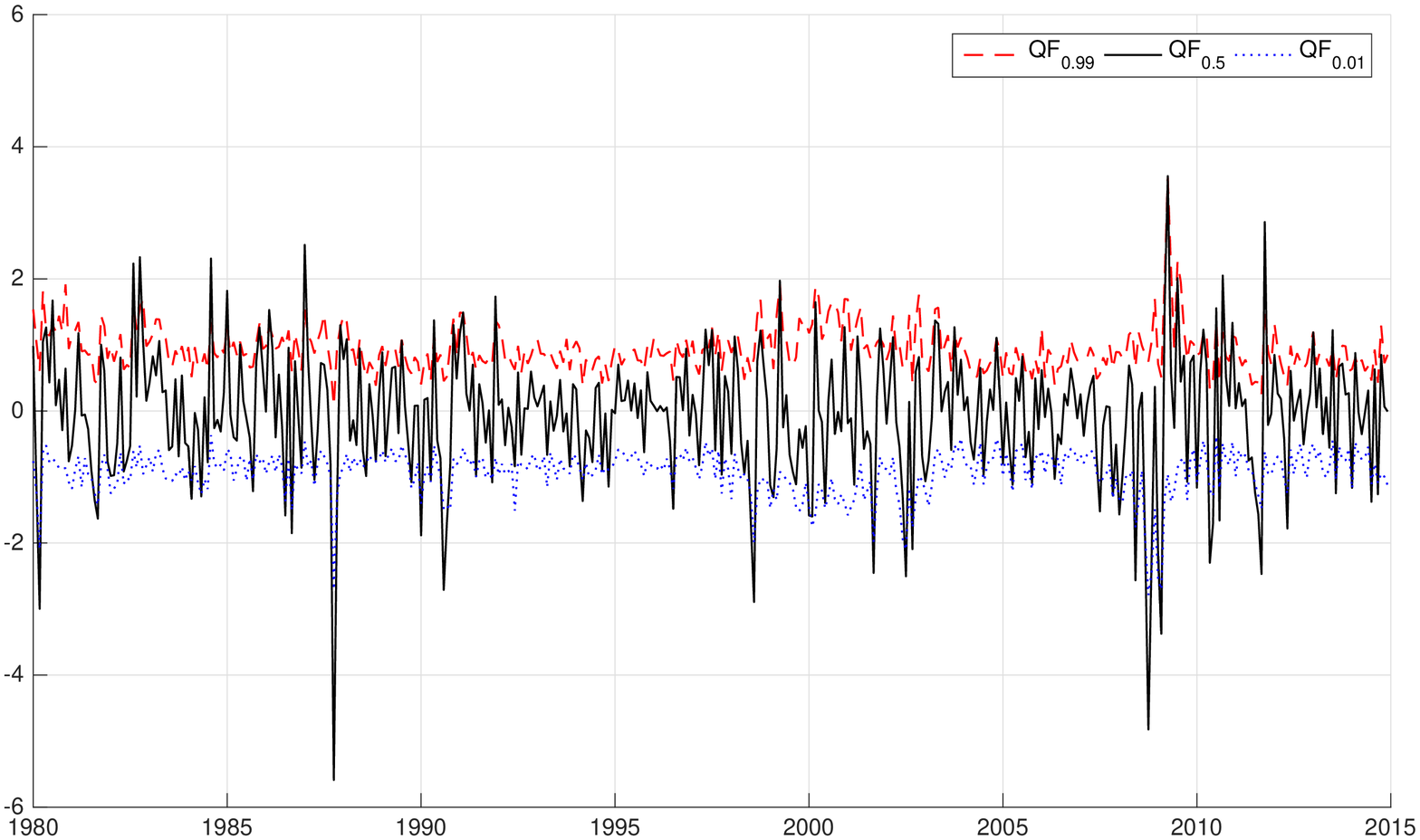}} 
\\
\subfloat{\includegraphics[width = 6in]{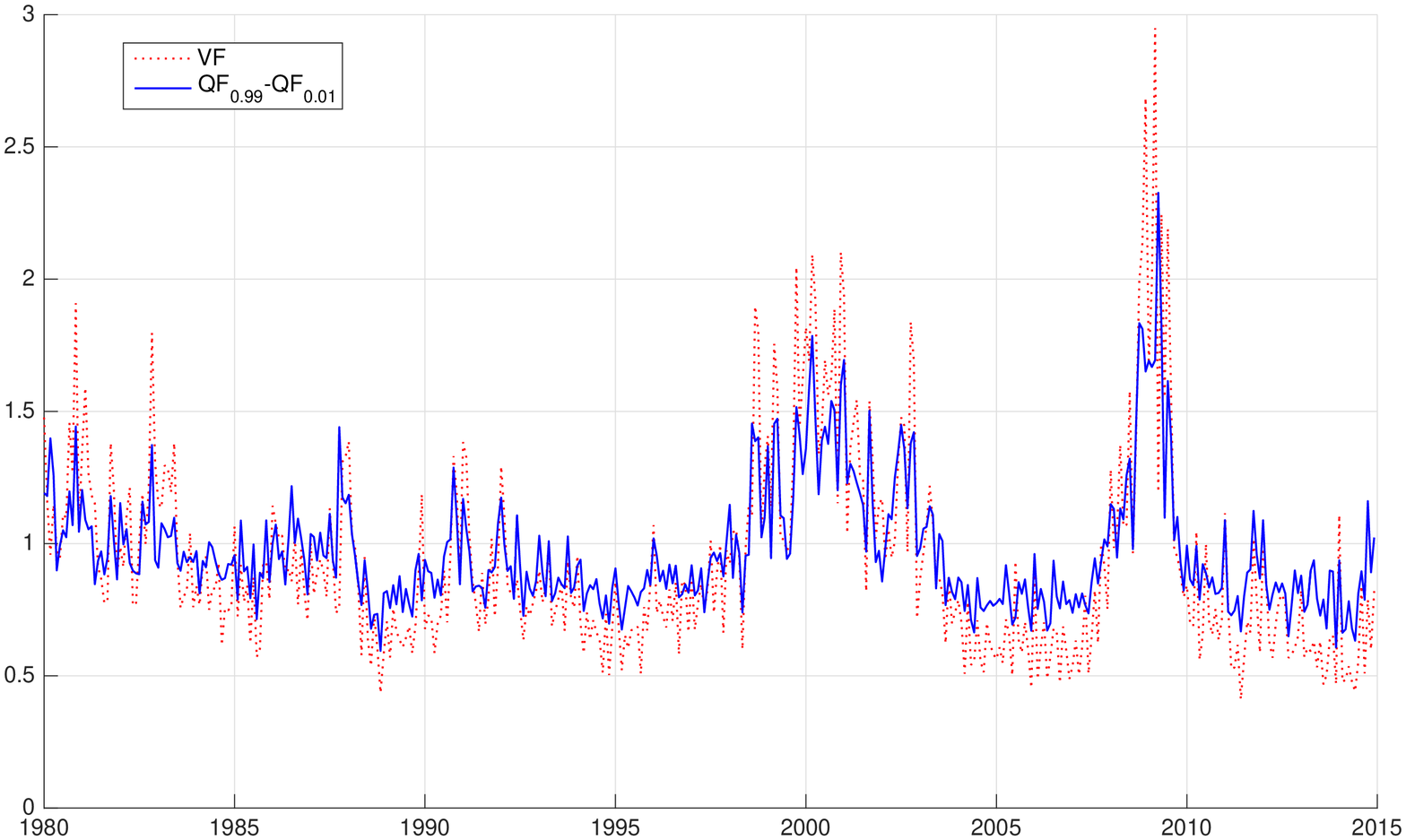}}
\label{some example}
{\centering
\begin{minipage}{6.5in}
    \small
    Note: The upper panel presents the QFA factors at $\tau=0.01, 0.09$ and the first QFA factor at $\tau=0.5$ from the financial datasets. The lower panel shows the interquantile range (blue line), defined as $\hat{F}_{QFA}^{0.99} - \hat{F}_{QFA}^{0.01} $, alongside the volatility factor (dotted red line) constructed by PCA-SQ. Both series are normalized to have unit length, and their correlation is 0.87.
    \end{minipage}
    }
\end{figure}
\par
}

\newpage
\begin{spacing}{0.8}
\bibliographystyle{chicago}
\bibliography{QFM}
\end{spacing}

\end{document}